\UseRawInputEncoding 
\documentclass[lettersize,journal]{IEEEtran}
\usepackage{amsmath,amsfonts,amssymb}
\usepackage{algorithmic}
\usepackage{algorithm}
\usepackage{array}
\usepackage[caption=false,font=normalsize,labelfont=sf,textfont=sf]{subfig}
\usepackage{textcomp}
\usepackage{stfloats}
\usepackage{url}
\usepackage{verbatim}
\usepackage{graphicx,amssymb,lineno}
\usepackage{cite}
\usepackage[usenames]{color}
\usepackage{float}
\usepackage{mathtools}
\usepackage{booktabs}

\usepackage{graphicx,graphics,color,epsfig,graphpap,rotate}
\usepackage{times, verbatim, epsfig, graphicx, latexsym, amsmath}
\usepackage{url}
\usepackage{CJK}

\begin{document}

\title{Energy Efficient Train-Ground mmWave Mobile Relay System for High Speed Railways}

\author{Lei~Wang,
        Bo~Ai,~\IEEEmembership{Fellow,~IEEE,}
        Yong~Niu,~\IEEEmembership{Member,~IEEE,}
        Zhangdui~Zhong,~\IEEEmembership{Fellow,~IEEE,}
        Shiwen~Mao,~\IEEEmembership{Fellow,~IEEE,}
        Ning~Wang,~\IEEEmembership{Member,~IEEE,}
        and~Zhu~Han,~\IEEEmembership{Fellow, IEEE}

\thanks{Copyright (c) 2015 IEEE. Personal use of this material is permitted. However, permission to use this material for any other purposes must be obtained from the IEEE by sending a request to pubs-permissions@ieee.org. This study was supported by the National Key Research and Development Program under Grant 2021YFB2900301; in part by National Key R\&D Program of China (2020YFB1806903); in part by the National Natural Science Foundation of China Grants 61801016, 61725101, 61961130391, and U1834210; in part by the State Key Laboratory of Rail Traffic Control and Safety (Contract No. RCS2021ZT009), Beijing Jiaotong University；and supported by the open research fund of National Mobile Communications Research Laboratory, Southeast University (No. 2021D09); in part by the Fundamental Research Funds for the Central Universities, China, under grant number 2022JBQY004 and 2022JBXT001; and supported by Frontiers Science Center for Smart High-speed Railway System; in part by the Fundamental Research Funds for the Central Universities 2020JBM089; in part by the Project of China Shenhua under Grant (GJNY-20-01-1). S. Mao's work is supported in part by the NSF Grant ECCS-1923717. Z. Han's work is partially supported by NSF CNS-2107216 and CNS-2128368. (\emph{Corresponding authors: B. Ai, Y. Niu.})}

\thanks{L.~Wang is with the State Key Laboratory of Rail Traffic Control and Safety, Beijing Jiaotong University, Beijing 100044, China, and also with Beijing Engineering Research Center of High-speed Railway Broadband Mobile Communications, Beijing Jiaotong University, Beijing 100044, China (email: lleiwang@bjtu.edu.cn).}
\thanks{B.~Ai is with the State Key Laboratory of Rail Traffic Control and Safety, Beijing Jiaotong University, Beijing 100044, China, and also with Peng Cheng Laboratory and Henan Joint International Research Laboratory of Intelligent Networking and Data Analysis, Zhengzhou University, Zhengzhou 450001, China (email: boai@bjtu.edu.cn).}
\thanks{Y.~Niu is with the State Key Laboratory of Rail Traffic Control and Safety, Beijing Jiaotong University, Beijing 100044, China, and also with the National Mobile Communications Research Laboratory, Southeast University, Nanjing 211189, China (email: niuy11@163.com).}
\thanks{Z.~Zhong is with the State Key Laboratory of Rail Traffic Control and Safety, Beijing Jiaotong University, Beijing 100044, China (e-mail: zhdzhong@bjtu.edu.cn).}
\thanks{S.~Mao is with the Department of Electrical and Computer Engineering, Auburn University, Auburn, AL 36949-5201 USA (email: smao@ieee.org).}
\thanks{N.~Wang is with the School of Information Engineering, Zhengzhou University, Zhengzhou, China, 450001 (email: ienwang@zzu.edu.cn).}
\thanks{Z. Han is with the Department of Electrical and Computer Engineering at the University of Houston, Houston, TX 77004 USA, and also with the Department of Computer Science and Engineering, Kyung Hee University, Seoul, South Korea, 446-701 (email: zhan2@uh.edu).}}

\maketitle

\begin{abstract}
The rapid development of high-speed railways (HSRs) puts forward high requirements on the corresponding communication system. Millimeter wave (mmWave) can be a promising solution due to its wide bandwidth, narrow beams, and rich spectrum resources. However, with the large number of antenna elements employed, energy-efficient solutions at mmWave frequencies are in great demand. Based on a mmWave HSR communication system with multiple mobile relays (MRs) on top of the train, a dynamic power-control scheme for train-ground communications is proposed. The scheme follows the regular movement characteristics of high-speed trains and considers three phases of train movement: the train enters the cell, all MRs are covered in the cell, and the train leaves the cell. The transmit power is further refined according to the number of MRs in the cell and the distance between the train and the remote radio head. By minimizing energy consumption under the constraints of the transmitted data and transmit power budget, the transmit power is allocated to multiple MRs through the multiplier punitive function-based algorithm. Comprehensive simulation results, where the velocity estimation error is taken into account, are provided to demonstrate the effectiveness of the proposed scheme over several baseline schemes.
\end{abstract}


\section{Introduction}
\IEEEPARstart{H}{igh-speed} railways (HSRs) are in high development due to its high mobility, great comfort, and high reliability. Compared to traditional means of transportation, HSR is changing how people travel and brings huge economic benefits while being convenient \cite{3}. The HSR network is rapidly expanding and will promote the development of various technologies, especially in the field of HSR communications \cite{4}. To be in line with future smart rail, HSR communication systems are expected to provide both train control services and mobile multimedia services for train passengers. With the help of smart technologies, we will not only see faster high-speed trains, but also high-speed data services for passengers, fully automated train operation and real-time monitoring in smart railway systems. Nevertheless, it is challenging to enable these high data rate required applications using current railway communication systems. The data rate of the most widely used global system for mobile communications for railways is at kb/s-level, and that of the long term evolution for railways is at Mb/s-level, which is still insufficient for many smart railway wireless communication services \cite{1}. As a result, millimeter wave (mmWave) communication systems attract significant interest.

MmWave can support multi-gigabit wireless data transmission, thus becoming a strong candidate for HSR communication systems to fulfill the increasing capacity requirements \cite{7}. However, it also brings about many new challenges. A major drawback is that mmWave communications suffer from blockage and increased path loss compared to communications in lower frequency bands \cite{12}. The propagation conditions at mmWave are more severer since mmWave signals cannot penetrate most solid materials \cite{17}. The solution to this problem is directional beamforming technique based on large-scale antenna arrays \cite{16}. Beamforming allows signals to be transmitted in a specific direction through the transmitter (TX) and receiver (RX) antennas, by which a highly directional transmission link is established.

Traditionally, base stations (BSs) have been the major power consumers in wireless communication networks, even in the absence of data transmission. This problem is even more severe in the mmWave band. Due to the high bandwidth, high volume of traffic and high transmit power, the energy consumption of a single mmWave BS is significantly higher than that of an existing single sub-6 GHz BS. Moreover, mmWave small BSs are usually deployed with high-power macro BSs in heterogeneous networks (HetNets) to increase system capacity. This means that massive traffic growth comes at the cost of huge energy consumption and a much larger carbon footprint. However, it is not desirable to increase system capacity through higher energy levels \cite{power-level}. Although mmWave technology can greatly improve the performance of HSR communication networks, these high data rate links also lead to increased device power consumption and a corresponding growth in system energy consumption. It is critical to overcome the energy consumption challenges of HSR mmWave communication networks due to the rising transmission rate demands.

Energy efficiency is a key performance metric for the fifth and future sixth generation communication systems, and has attracted extensive attention from both academia and industry \cite{greenj}. The design of future wireless communication networks should take energy efficiency into account and meet more stringent energy efficiency requirements. HSR is widely recognized as a green transportation that requires an organic combination of energy efficiency and functional design \cite{1}. It is of great practical importance to study the energy efficiency problem of train-ground mmWave communications for HSRs.

\subsection{Related Work}\label{S1-1}

A significant amount of work is focused on the resource allocation to improve the energy efficiency of wireless communication systems \cite{ra1,eh1,ra2,ra3,ra4,eh4,ra5,ra6}. With the goal of maximizing the energy efficiency of orthogonal frequency division multiple access HetNets employing wireless backhaul, Ref. \cite{ra1} designs power and bandwidth allocation schemes by decoupling the joint optimization problem into two convex optimization problems. Zhang \emph{et al}. in \cite{eh1} consider a HetNet powered by harvested energy, on-grid energy or both, and derive a closed-form expression for the power saving gain. Traffic offloading schemes for a single SBS and multiple SBS scenarios are developed to minimize on-grid energy consumption under the quality of service (QoS) constraint. A multi-objective optimization problem subject to channel, time slot, transmission power and QoS constraints is formulated in \cite{ra2} to exploit the trade-off between energy efficiency and spectral efficiency in device-to-device (D2D) communications supporting energy harvesting. Energy and task allocation in wireless-powered mobile edge computing networks are also solved by convex optimization techniques in \cite{ra3}. In particular, the authors consider randomly arriving tasks and situations where future channel state information is unknown. A game theory-based approach is proposed for sub-channel and power allocation in ultra dense networks where long term evolution (LTE) and Wireless Fidelity (WiFi) coexist \cite{ra4}. Ref. \cite{eh4} studies the energy scheduling problem of D2D communication with energy harvesting capability, especially considering the energy consumption of the device to process data. In recent years, deep reinforcement learning approaches have also been used to solve energy efficiency maximization problems \cite{ra5,ra6}.

The energy consumption of mmWave communication systems is of particular concern due to the high frequency bands and the large number of antenna elements. Several optimization schemes have been proposed to achieve energy efficient  mmWave communications \cite{related2,related,32,31}. For mmWave cell-free systems, Ref. \cite{related2} selects several main paths in the angle domain and performs information feedback and transmission power allocation on these paths. Ref. \cite{related} proposes to utilize the energy recovered from radio frequency signals and coordinate data transmission through multiple BSs to improve the energy efficiency of ultra-dense HetNets with mmWave massive multiple-input multiple-output (MIMO). In~\cite{32}, the authors focus on the design of analog beamformers in mmWave multi-input single-output systems and propose a low-complexity solution under power constraints. Digital and hybrid mmWave beams are also designed in~\cite{31} through a hybrid mapping algorithm to maintain the dynamic balance between the energy efficiency and beam ripples.

Recently, mmWave HSR communications have been extensively studied \cite{3GPP,cn,28,29,channel}. The throughput performance of the HSR communication system with a two-hop architecture has been shown to outperform the performance of direct communication between BSs and passengers, and deploying multiple independent mobile relays (MRs) is superior to deploying a single MR \cite{3GPP}. In this two-hop network, the type of MR can be either amplify-and-forward or decode-and-forward, and there is a trade-off between the performance and cost of the MR. 5G New Radio (NR) is believed to further enhance the performance of HSR communication systems. Ref.~\cite{cn} describes the performance requirements for deploying 5G NR systems in HSR scenarios and provides the physical layer design and initial access mechanism to support high-speed mobile scenarios. Conventional beam alignment methods may introduce large angular offsets in HSR systems, the authors in~\cite{28} propose a fast initial access scheme by taking advantage of learning results from historical beam training process. Moreover, a network structure which utilizes low frequency bands to improve the performance of mmWave frequencies is applied to guarantee the robustness of the entire network. Xu \emph{et al}. in \cite{29} consider a practical signal propagation environment, and propose a channel tracking scheme based on angle information, while a hybrid beamforming scheme is also designed to reduce overhead. A channel model for mmWave HSR systems is developed in \cite{channel}, which is a three-dimensional model that captures channel non-stationarity in time, space and frequency.

Notably, most of these prior works do not consider the energy efficiency optimization for HSR communications. Although there have been some studies on energy efficiency related problems, there are not many discussions on energy-efficient problem in the unique and challenging HSR scenario. Due to the fast moving speed of high-speed trains, HSR communications suffer from severe Doppler shift and penetration loss, and these problems are more severe at mmWave frequencies. In addition, frequent handovers are performed in HSR systems, which is launched by a large number of user equipment almost simultaneously and has to be completed in a short period of time \cite{handover}. To address these issues, the third Generation Partnership Project (3GPP) has adopted a two-hop architecture in HSR, where data from passengers to BSs is forwarded by roof MRs \cite{3GPP}. High-speed trains provide a lot of space for large-scale antennas, and hence roof-top MR can also provide high-speed data transmission services for HSR passengers with the help of high frequency bands such as mmWave \cite{zhangjiayi}. Moreover, multiple MRs are expected to further enhance system throughput. However, different from single MR, multi-MRs consumes more energy. Providing maximum transmit power to each MR results in the system operating at a much lower efficiency than its optimal capability, and the power consumed by the system is not fully utilized and is greatly wasted. Therefore, how to achieve dynamic power allocation among multiple MRs to save the energy of HSR communication systems while meeting the data transmission requirements is a key issue.

\subsection{Contributions and Organization}\label{S1-2}

In this paper, we consider the train-ground mmWave communication for HSRs, where an mmWave remote radio head (RRH) can serve multiple MRs installed on the train simultaneously \cite{mu}. Directional beamforming is employed at the TX and RX, to compensate for the path loss at higher frequencies. Then an energy efficiency optimization problem is formulated and a power allocation scheme based on the multiplier punitive function algorithm is developed. The scheme considers three phases of train movement: (i) the train enters the cell, (ii) all the MRs are covered in the cell, and (iii) the train leaves the cell. The transmit power of the second phase is dynamically adjusted according to the distance between the train and the RRH, and the transmit power of the other two phases is allocated according to the number of MRs in the cell. Main contributions are as follows.

\begin{itemize}
	\item This paper investigates the energy efficiency problem of mmWave multi-MR HSR systems, where more than one MR is mounted on the roof of the train. A dynamic power allocation scheme is designed to support a number of RRH-MR links simultaneously. In particular, the process of the train entering or leaving the cell is divided into several stages depending on the number of MRs in the cell, and when all MRs are in the cell, the power is allocated sequentially in several consecutive location bins. 
	
	\item An optimization problem is formulated to achieve high energy efficiency of train-ground communications. The optimization problem imposes a strict constraint on the transmit power of the system, that is, the sum of the transmit powers of multiple MRs is less than the transmit power budget. In addition, energy consumption is minimized under the constraint that the total transmitted data is greater than the threshold. 
	
	\item The formulated optimization problem is a multivariate non-convex non-linear problem, which is difficult to obtain the optimal power allocation results. A multiplier punitive function-based algorithm is proposed to realize power allocation of different MRs during the train movement. Simulations verify that the optimized power allocation scheme can improve energy efficiency of the system. Moreover, the impact of errors in estimating train speed is also analyzed.
\end{itemize}

The remainder of this paper is organized as follows. Section~\ref{S2} describes the system model and formulates the energy efficiency maximization problem into a non-convex non-linear minimization problem. Section~\ref{S3} proposes a multiplier punitive function-based algorithm to solve the optimization problem. Section~\ref{S4} presents performance evaluation and explores the effect of speed error. Finally, Section~\ref{S5} concludes this paper.

\section{System Model and Problem Formulation}\label{S2}

\subsection{System Model}\label{S2-1}

\begin{figure}[!t]
\begin{center}
\includegraphics*[width=2.5in]{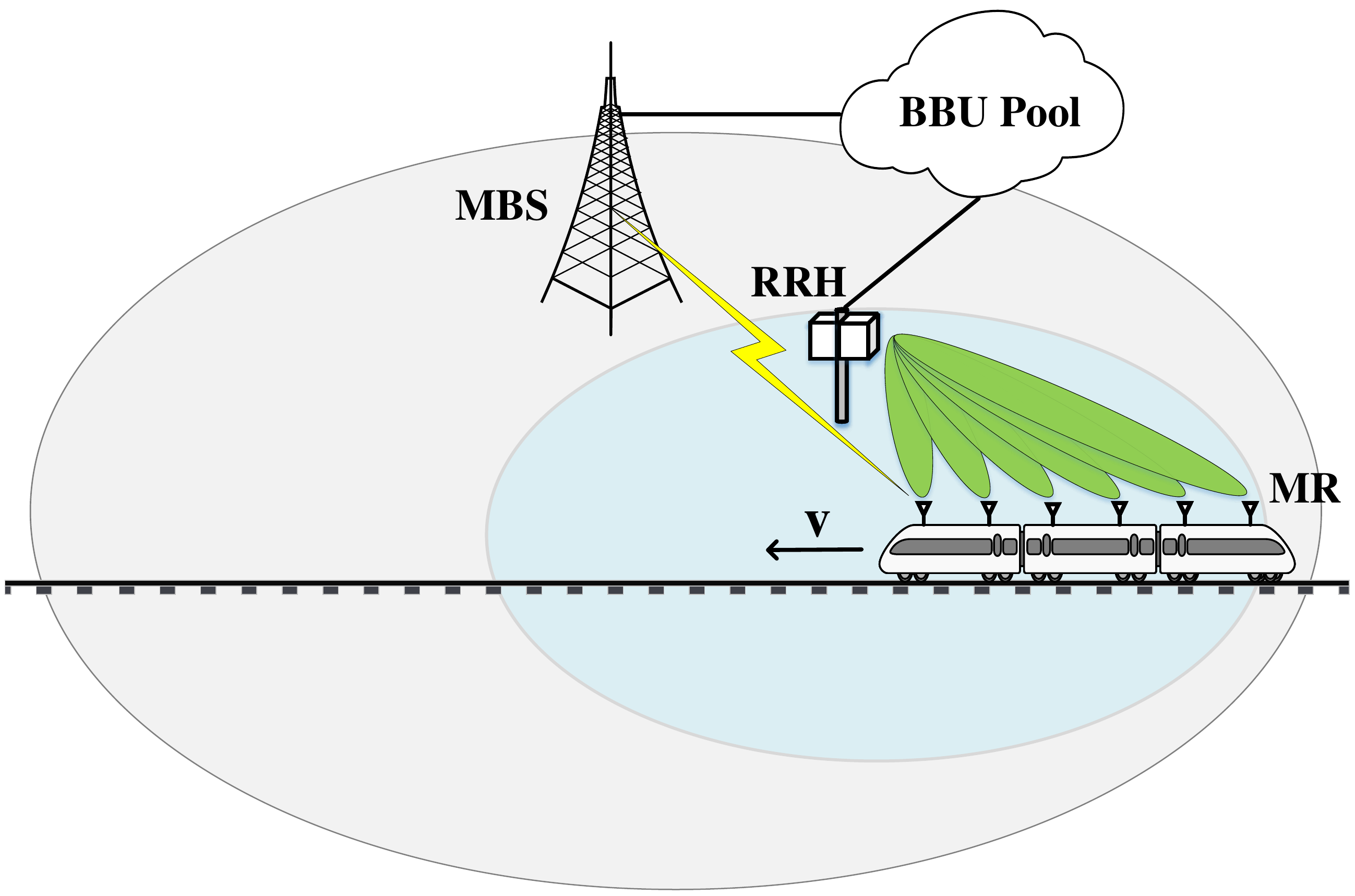}
\end{center}
\caption{A control/user-plane splitting network for HSR communications.}
\label{network}
\end{figure}

To meet the high-speed transmission demands of HSR passengers, a control/user-plane splitting network architecture is adopted in this paper \cite{1}, as shown in Fig. \ref{network}. Control signaling is carried on the control-plane, and data information from passengers is offloaded to the track-side mmWave RRH. The network is also a HetNet deployment, with the mmWave RRH deployed within the coverage of macro cell for high data rate transmission, and MBS operating in lower frequency bands for coverage and reliability. Due to the high cost, it is impractical to deploy continuous mmWave RRHs. This separation of user plane and control plane allows flexible deployment of mmWave RRHs to improve transmission rates in small areas while maintaining coverage performance. 

To provide reliable connectivity for passengers, multiple MRs are deployed on top of the train, forming a two-hop architecture. MRs are mainly used for receiving and forwarding data, and all communications between RRHs and passengers are completed through MRs, thus avoiding severe train body penetration loss. Compared to a single MR, multiple MRs take advantage of space diversity gain to further improve system throughput. In the first hop, MRs establish connections with the RRH through radio access links, which can be both at frequency bands below 6 GHz and mmWave frequencies. In the second hop, MRs serve onboard passengers through the access points installed inside each carriage, where MRs and access points are connected by the wired cable or fiber. At the access links, various well-developed radio access technologies are enabled, like LTE and WiFi \cite{gao2020Efficient}. Another advantage of using this two-hop structure is that the data from passengers first arrives at MRs and then is forwarded to the low-frequency BSs via MRs, avoiding frequent handover when passengers directly communicate with BSs. Since communication requests and data information from passengers are aggregated at MRs, we only consider the RRH-MR links that support the train-ground mmWave communications.  

Assume that all transceivers are equipped with adaptive antenna arrays and a beam switching method is employed to achieve effective beam alignment. Train position information from the train control system and train velocity measured by the train itself is exploited to compute the real-time train position to further assist beam switching. Then train position can be predicted by
\begin{equation}
x(t)=v(t-t_0)+x_0, \label{eq1}
\end{equation}
where $x(t)$ is the train position at time $t$, $v$ is the train speed, $x_0$ is the feedback location from the train control system at time $t_0$. In practice, train speed information may not be perfect, and speed estimation errors may exist. The effect of velocity estimation error will be discussed in Section \ref{S4-3}.

Let $d_0$ be the distance between the RRH and the rail, and $d_l$ is the coverage width of the RRH. Suppose the train has $M$ MRs, and the distance between adjacent MRs is $d_{MR}$ which satisfies ${d_l} > \left( {M - 1} \right){d_{MR}}$, i.e., the movement of the train consists of three processes: the train enters the cell, all $M$ relays of the train are covered in the cell, and the train leaves the cell. 

The mmWave channel model in dB is expressed as
\begin{equation}
{h_i}(t) =  - P{L_i}\left( t \right) - \xi  - \gamma \left( t \right), \label{h}
\end{equation}
where ${h_i}(t)$ is the channel coefficient between the $i$th MR and the RRH at time $t$, the path loss is denoted by $P{L_i}\left( t \right)$ and is given by $P{L_i}\left( t \right) = 10n{\log _{10}}\left( {{{4\pi {d_i}\left( t \right)} \mathord{\left/
 {\vphantom {{4\pi {d_i}\left( t \right)} \lambda }} \right.
 \kern-\nulldelimiterspace} \lambda }} \right)$, where $n$ denotes the path loss exponent, $\lambda$ denotes the carrier wavelength, and ${d_i}\left( t \right)$ denotes the distance between the track-side RRH and the MR $i$. $\xi$ is the shadowing margin, and $\gamma \left( t \right)$ denotes the scaling fading factor incorporating Rician fading whose envelope is distributed as 
\begin{equation}
f(r) ={r\exp \left( { - {{\left( {{r^2} + {A^2}} \right)} \mathord{\left/{\vphantom {{\left( {{r^2} + {A^2}} \right)} {2{\sigma ^2}}}} \right.\kern-\nulldelimiterspace} {2{\sigma ^2}}}} \right) \cdot {I_0}\left( {{{rA} \mathord{\left/{\vphantom {{rA} {{\sigma ^2}}}} \right.\kern-\nulldelimiterspace} {{\sigma ^2}}}} \right)}/{{\sigma ^2}}.
\end{equation}

To mitigate the negative impact of Doppler shift on HSR mmWave communication systems, a machine learning-based Doppler shift estimation method is adopted \cite{justification_ML}. With the powerful processing capability of MRs, a Doppler shift estimation model, which is a trained neural network, is deployed at one of the MRs. 

Considering the regularity of train movement, the proposed model constructs the mapping relationship from reference signal received power (RSRP) values to Doppler shift by the pattern of RSRP. Specifically, the input set of the model consists of $2L+1$  RSRP values around position $x_a$, which is given by
\begin{equation}
\begin{aligned}
s\left( {{x_a}} \right) = &\left\lbrace r\left( {{x_{a - L}}} \right), \cdots ,r\left( {{x_{a - 1}}} \right),r\left( {{x_a}} \right),r\left( {{x_{a + 1}}} \right), \cdots ,\right. \\
& \left. r\left( {{x_{a + L}}} \right) \right\rbrace,
\end{aligned}
\end{equation}
where $s\left( {{x_a}} \right)$ denotes the input set at position $x_a$, $r\left( {{x_j}} \right),j = a - L, \cdots ,a, \cdots ,a + L$ denotes the RSRP value at position $x_j$ and is given by \cite{34}
\begin{equation}
r\left( {{x_j}} \right) = {P_T} + {G_{tx}} + {G_{rx}} - \xi  + 10n{\log _{10}}\left( {{\lambda  \mathord{\left/
 {\vphantom {\lambda  {4\pi d\left( {{x_j}} \right)}}} \right.
 \kern-\nulldelimiterspace} {4\pi d\left( {{x_j}} \right)}}} \right) - \gamma \left( t \right),
\end{equation}
where $P_T$ is the transmit power budget; $G_{tx}$ and $G_{rx}$ denote the TX and RX antenna gains, respectively; $d\left(x_j\right)$ is the distance of the MR from the RRH at position $x_j$. Since different locations may correspond to the same RSRP value, if a single RSRP value is directly input to the model, it will result in the same Doppler shift value at different locations. Multiple input values avoid this one-to-many mapping problem. 

The RSRP values are measured every $x_s$ meters, that is, every ${{{x_s}} \mathord{\left/{\vphantom {{{x_s}} v}} \right.\kern-\nulldelimiterspace} v}$ seconds, to generate the training data of the model. By training a neural network, a Doppler shift estimator is obtained. However, the model can only estimate the Doppler shift at the velocity parameter for which the model was trained. Estimates at other speeds require extensive training of the model at different speeds. To address this, the output of the model is redefined as the relative Doppler shift, denoted by $f_d^{rel}\left( x \right)$, and is given by
\begin{equation}
f_d^{rel}\left( x \right) = \frac{{{f_d}\left( x \right)}}{{{f_{d,\max }}}},
\end{equation}
where ${f_d}\left( x \right)$ is the Doppler shift at position $x$, $f_{d,\max }$ is the maximum Doppler shift which is computed by ${f_{d,\max }} = {{{f_c}v} \mathord{\left/{\vphantom {{{f_c}v} c}} \right.\kern-\nulldelimiterspace} c}$ where $f_c$ is the carrier frequency and $c$ is the speed of light. 

Finally, the Doppler shift can be estimated by 
\begin{equation}
{\hat f_d}\left( x \right) = \frac{{f_d^{rel}\left( x \right){f_c}v}}{c}.
\end{equation}

\subsection{Problem Formulation}\label{S2-2}

The energy efficiency optimization for HSR communications can be formulated as
\begin{equation}
\max\ EE=\frac{D}{E},
\end{equation}
where $EE$ is the energy efficiency, $D$ is the total transmitted data, and $E$ denotes the total energy consumption.

Consider the entire process from the first MR of the train entering the cell to the last MR leaving the cell, which consists of three stages: (i) MRs of the train enter the cell sequentially, (ii) all the $M$ MRs are covered in the cell, and (iii) MRs leaves the cell in turn. The first stage can be further divided into $M-1$ parts according to the number of MRs that have entered the cell; so is the third stage. Moreover, the distance travelled by the train in the second stage is divided into $N$ consecutive location bins of length ${{\left[ {{d_l} - \left( {M - 1} \right){d_{MR}}} \right]} \mathord{\left/ {\vphantom {{\left[ {{d_l} - \left( {M - 1} \right){d_{MR}}} \right]} N}} \right. \kern-\nulldelimiterspace} N}$.

Let the time when the first MR reaches the cell edge be time $0$, i.e., $t_0=0$, and then the running time of the train can be divided into $2M+N-2$ segments according to the above approach. 

Let the time instances for the first MR to reach the endpoint of each segment be $t_i$, given by
\begin{equation}
	{t_i} = \left\{ \begin{array}{l}
		\dfrac{{i{d_{MR}}}}{v},i \in \left[ {{\rm{1,2,}} \ldots ,M - 1} \right]\vspace{1.5ex},\\
		\dfrac{{{a_i}{d_{MR}} + {b_i}{d_l}}}{{v \cdot N}},i \in \left[ {M,M + 1, \ldots ,M + N - 1} \right]\vspace{1.5ex},\\
		\dfrac{{{c_i}{d_{MR}} + {d_l}}}{v},i \in \left[ {M + N, \ldots ,2M + N - 2} \right],
	\end{array} \right.
\end{equation}
where
\begin{equation}
\begin{aligned}
{a_i}& = \left( {N + M - i - 1} \right)\left( {M - 1} \right),\\
{b_i}&= i - M + 1,\\
{c_i}&= \left( {i - M - N + 1} \right).
\end{aligned}
\end{equation}

When $t \in \left[ {0,{t_{M - 1}}} \right)$, the train enters the cell during this interval, corresponding to the first stage. When $t \in \left[ {{t_{M - 1}},{t_{M + N }}} \right)$, all the $M$ relays are covered in the cell, corresponding to the second stage. And when $t \in \left[ {{t_{M + N }},{t_{2M + N - 2}}} \right)$, the train leaves the cell, corresponding to the third stage. In addition, when $t \in \left[ {{t_{i - 1}},{t_i}} \right),i \in \left[ {{\rm{1,2,}} \ldots ,M - 1}\right]$, there are $i$ MRs covered in the cell; when $t \in \left[ {{t_{i - 1}},{t_i}} \right),i \in \left[ {M + N, \ldots ,2M + N - 2} \right]$, there are $2M+N-i-1$ MRs covered in the cell.

Since the length of location bins and the distance between MRs are both short, assume that the transmit power allocated to each MR during each interval $\left[ {{t_{i - 1}},{t_i}} \right)$ is constant.  Let ${\boldsymbol{P}} = \left[ {{P_{i,j}}} \right]$ denote a $M \times \left( {2M + N - 2} \right)$ transmit power allocation matrix, where ${P_{i,j}}$ denotes the transmit power allocated to the $i$th MR during $\left[ {{t_{j - 1}},{t_j}} \right)$. If the MR has not entered the cell or has left the cell, there is no transmit power allocated to it.

The total energy consumption $E$ of the train-ground mmWave communications can be calculated as
\begin{equation}\label{eq:E}
	E = {\boldsymbol{T}}{{\boldsymbol{P}}^T}{{\boldsymbol{I}}} ,
\end{equation}
where ${\left(  \cdot  \right)^T}$ denotes the transpose operator. ${{\boldsymbol{I}}}$ is a $M \times 1$ matrix with elements 1.  ${\boldsymbol{T}}$ is a time matrix and is expressed as
\begin{equation}
\begin{aligned}
		{\boldsymbol{T}} =  \left[\right.& \frac{{{d_{MR}}}}{v},  \ldots  ,\frac{{{d_{MR}}}}{v},\frac{{{d_l} - \left( {M - 1} \right){d_{MR}}}}{{v \cdot N}}, \ldots ,\\
		&\left. \frac{{{d_l} - \left( {M - 1} \right){d_{MR}}}}{{v \cdot N}},\frac{{{d_{MR}}}}{v}, \ldots ,\frac{{{d_{MR}}}}{v} \right],
\end{aligned}
\end{equation}
where ${\boldsymbol{T}}$ consists of $2M+N-2$ elements, the first $M-1$ and the last $M-1$ elements are ${d_{MR}} / v$, and the middle $N$ elements are $\left({d_l} - \left( {M - 1} \right){d_{MR}}\right)/v / N$.

The received power of MR $i$ during $\left[ {{t_{j - 1}},{t_j}} \right)$, denoted by $P_{rx,i,j}$, is given by
\begin{equation}
	P_{rx,i,j}(t)= P_{i,j} +G_{tx}+G_{rx} - \xi + 10n{\log _{10}}\frac{\lambda }{{4\pi d_i(t)}}-\gamma \left( t \right) [dBm].
\end{equation}

The received SNR of MR $i$ at time $t$, denoted by ${\Gamma _{i}\left(t\right)}$, is obtained as 
\begin{equation}
	{\Gamma _{i}\left(t\right)} = {P_{i,j}} + 10n{\log _{10}}\frac{\lambda }{{4\pi d_i\left( t \right)}} + \gamma \left( t \right)+ {C_{i}},
\end{equation}
where 
\begin{equation}
	{C_{i}} = {G_{tx}} + {G_{rx}} - \xi - P_{noise}^{dBm},
\end{equation}
\begin{equation}
	P_{noise}=-174+10\log_{10}{B}+\mathrm{NF}\ [dBm],
\end{equation}
where $P_{noise}$ is the noise power, $B$ is the system bandwidth, and NF is the noise figure.

According to the Shannon capacity formula, the amount of transmitted data of MR $i$ during $\left[ {{t_{j - 1}},{t_j}} \right)$, denoted by ${D_{i,\left[ {{t_{j - 1}},{t_j}} \right)}}$, is given by
\begin{equation}
	{D_{i,\left[ {{t_{j - 1}},{t_j}} \right)}} = \int_{{t_{j - 1}}}^{{t_j}} {{{\log }_2}\left( {1 + {{10}^{{{{\Gamma _i}\left( t \right)} \mathord{\left/{\vphantom {{{\Gamma _i}\left( t \right)} {10}}} \right.\kern-\nulldelimiterspace} {10}}}}} \right)dt} .
\end{equation}

Then the total amount of transmitted data $D$ can be computed as
\begin{equation}
	D = \sum\limits_{i = 1}^M {\sum\limits_{j = i}^{i + M + N - 2} {{D_{i,\left[ {{t_{j - 1}},{t_j}} \right)}}} }. \label{eq:D}
\end{equation}

Equations \eqref{eq:E}-\eqref{eq:D} provide a model for the energy efficiency $EE$. Since $EE = D{\rm{/}}E$, we optimize $EE$ through minimizing $E$ under the constraint that $D$ meets the preset threshold $D_{min}$. Then the optimization problem can be modeled as
\begin{equation}
	\begin{aligned}
		\min\quad& E  \\
		\rm{s.t.}\quad &D\ge{D_{min}},
	\end{aligned}
\end{equation}
where $D_{min}$ is the minimum requirement for the amount of transmitted data.

In addition to the constraint on the total amount of transmitted data, the sum of the transmit powers allocated to the MRs in each period must be less than the transmit power budget $P_T$.

Finally, the energy efficiency optimization problem for HSR communications can be modeled as
\begin{subequations}
	\begin{align}
		\mathop {\min }\limits_{\boldsymbol{P}}\quad&E \nonumber\\
		\rm{s.t.}\quad&D \ge {D_{min}},\label{e1}\\
		&{\left\| \boldsymbol{P} \right\|_1} \le {P_T},\label{e2}
	\end{align}\label{E}
\end{subequations}
where ${\left\|  \cdot  \right\|_1}$ denotes the maximum absolute column sum norm and is given by ${\left\| \boldsymbol{A} \right\|_1} = \mathop {\max }\limits_{j = 1,2, \ldots ,n} \sum\limits_{i = 1}^m {\left| {{A_{i,j}}} \right|} $. Accordingly, \eqref{e2} implies that 
\begin{equation}
\sum\limits_{i = 1}^M {{P_{i,j}}}  \le {P_T},\forall j \in \left[ {1,2, \ldots ,2M + N - 2} \right],
\end{equation}
That is, the problem \eqref{E} actually has $2M + N - 1$ constraints with $M \times \left( {M + N - 1} \right)$ optimization variables. Since the constraint \eqref{e1} contains non-convex SNR parts, the problem is a non-convex non-linear optimization problem.

\section{Proposed Power-Control Algorithm}\label{S3}

Due to the non-linearity of the constraints \eqref{e1}, it is not possible to transform this problem into an unconstrained problem by the elimination method. One approach is to form an auxiliary function from the objective function and the constraint functions, transforming the original constrained problem into an unconstrained problem that minimizes the auxiliary function. 

The multiplier punitive function method combines the advantages of the Lagrange multiplier method and the penalty function method, that is, fast convergence and the ability to transform a constrained problem into an unconstrained one, and is therefore suitable for the above problem \eqref{E}. However, directly solving the inequality constrained optimization problem requires the introduction of new variables to transform inequality constraints into equality constraints, which will undoubtedly greatly increase the complexity of the algorithm. Since the optimal solution can be obtained from the Karush-Kuhn-Tucker (KKT) conditions, we first analyze the KKT conditions here.

The Lagrange function for problem (\ref{E}) is given by
\begin{equation}
	\mathcal{L}{\left( {{\boldsymbol{P}};{\boldsymbol{\lambda }}} \right)} = E - {\lambda _1}\left( {D - {D_{\min }}} \right) + {\lambda _2}\left( {{{\left\| \boldsymbol{P} \right\|}_1} - {P_T}} \right),
\end{equation}
where ${\lambda_i}(i \in [1,2] )$ are Lagrange multipliers.

Construct corresponding KKT conditions as follows.
\begin{equation}\label{kkt}
	\left\{ \begin{array}{l}
		{\nabla _p}\mathcal{L}\left( {{\boldsymbol{P}};{\boldsymbol{\lambda }}} \right) = 0\vspace{1.3ex},\\
		{\lambda _1}\left( {D - {D_{\min }}} \right)=0 \vspace{1.3ex},\\
		{\lambda _2}\left( {{{\left\| \boldsymbol{P} \right\|}_1} - {P_T}} \right)=0 \vspace{1.3ex},\\
		D - {D_{\min }} \ge 0 \vspace{1.3ex},\\
		{\left\| \boldsymbol{P} \right\|_1} - {P_T} \le 0\vspace{1.3ex},\\
		{\lambda_i} \ge 0,i \in [1,2].
	\end{array} \right.
\end{equation}

Due to the first constraint \eqref{e1} of problem \eqref{E} involves exponential and logarithmic operations, it is difficult to solve the KKT constraints directly. On the other hand, through analysis and calculation of \eqref{kkt}, it is found that the solution to the optimization problem can be obtained when the constraints are all equal, and thus, the inequality constrained optimization problem (\ref{E}) is transformed as follows
\begin{subequations}\label{eq:36}
	\begin{align}
		\mathop {\min }\limits_{\bf{P}}\quad&E \nonumber\\
		\rm{s.t.}\quad&D = {D_{min}},\\
		&\sum\limits_{i = 1}^M {{P_{i,j}}} = {P_T},\forall j \in \left[ {1,2, \cdots ,2M + N - 2} \right]\label{36-2}.
	\end{align}
\end{subequations}

The multiplier punitive method forms an augmented Lagrangian function to transform the constrained optimization into an unconstrained optimization, and algorithms for solving the unconstrained optimization problem can be directly used to solve the original problem. Since the method does not involve constraints at each iteration, it is suitable for solving non-linear constrained optimization problems. The augmented Lagrange function of problem~\eqref{eq:36} is given by
\begin{equation}
	\begin{aligned}
	\phi \left( {\boldsymbol{P},\boldsymbol{\lambda} ,\sigma } \right) &= E - {\lambda _1}\left( {D - {D_{\min }}} \right) \\
	& - \sum\limits_{j = 1}^{2M + N-2 } {{\lambda _{j+1}}\left( {\sum\limits_{i = 1}^M {{P_{i,j}}}  - {P_T}} \right)}  \\
	&+ \sigma {\left( {D - {D_{\min }}} \right)^2} \\
	&+\sigma {\sum\limits_{j = 1}^{2M + N - 2} {\left( {\sum\limits_{i = 1}^M {{P_{i,j}}}  - {P_T}} \right)} ^2},
	\end{aligned}
\end{equation}
where 
$\sigma$ is a penalty factor.

\begin{algorithm}[t!]
	\caption{The Multiplier Punitive Function-based Power Control Algorithm}
	\label{alg:1}
	\textbf{Initialization}:\ pick ${\sigma ^{\left(0\right)} }> 0$, ${{\boldsymbol{\lambda }}^{\left(0\right)}} = {\boldsymbol{0}}$, stepsize $\alpha$, growth factor $\gamma  > 1$, allowable error $\varepsilon > 0$, the number of cycles $n$ ;\\
	\textbf{Output}:\ ${\boldsymbol{\bar P}}$ ;
	\begin{algorithmic}[1]
		\STATE Set $k=0$ ;
		\WHILE {$k\le n$}
		\STATE Set $i=1$ and pick initial point ${{\boldsymbol{P}}^{\left(1\right)}}$;
		\WHILE {true}
		\STATE Compute descent direction ${\boldsymbol{d}^{\left( i \right)}} =  - \nabla \phi \left( {\boldsymbol{P^{\left( i \right)}},\boldsymbol{\lambda^{{\left( k \right)}}} ,\sigma^{{\left( k \right)}} } \right) $;
		\IF 	{$\left\| {{\boldsymbol{d}^{\left( i \right)}}} \right\| \le \varepsilon $} \vspace{1.1ex}
		\STATE Break ;
		\ELSE 
		\STATE ${{\boldsymbol{P}}^{\left( {i + 1} \right)}} = {{\boldsymbol{P}}^{\left( i \right)}} + \alpha {\boldsymbol{d}^{\left( i \right)}}$, $i=i+1$;
		\ENDIF
        \ENDWHILE
		\STATE Set ${{\boldsymbol{P}}^{\left(k\right)}}={{\boldsymbol{P}}^{\left(i\right)}}$;
		\IF {$\max \left\{ {{h_j}\left( {{\boldsymbol{P}}^{\left( k \right)}} \right)\left| {j = 1, \cdots ,2M + N - 1} \right.} \right\} \le \varepsilon$}
		\STATE Break ;
		\ELSIF {${\left\| {{\boldsymbol{h}}\left( {\boldsymbol{P}}^{\left( k \right)}\right)} \right\|_\infty } \ge {\left\| {\boldsymbol{h}\left( {{\boldsymbol{P}}^{\left( k-1 \right)}} \right)} \right\|_\infty }$}\vspace{1.1ex}
		\STATE ${\sigma ^{\left(k+1\right)}} = \gamma {\sigma^{\left(k\right)}}$, ${{\boldsymbol{\lambda }}^{\left(k+1\right)}} = {{\boldsymbol{\lambda }}^{\left(k\right)}}$, $k=k+1$ ;\vspace{1.1ex}
		\ELSIF {$\sigma^{\left(k\right)} > {\sigma ^{\left(k-1\right)}}$ or ${\left\| {{\boldsymbol{h}}\left( {\boldsymbol{P}}^{\left( k \right)}\right)} \right\|_\infty } \le \dfrac{1}{4}{\left\| {\boldsymbol{h}\left( {{\boldsymbol{P}}^{\left( k-1 \right)}} \right)} \right\|_\infty }$}\vspace{1.1ex}
		\STATE $\sigma ^{\left(k+1\right)} = {\sigma^{\left(k\right)}}$, ${\boldsymbol{\lambda } ^{\left( {k + 1} \right)}} = {\boldsymbol{\lambda } ^{\left( k \right)}} - 2\sigma^{\left(k\right)} \boldsymbol{h}\left( {{{\boldsymbol{P}}^{\left( k \right)}}} \right)$, $k=k+1$ ;
		\ELSE
		\STATE ${\sigma ^{\left(k+1\right)}} = \gamma {\sigma^{\left(k\right)}}$, ${\boldsymbol{\lambda } ^{\left( {k + 1} \right)}} = {\boldsymbol{\lambda } ^{\left( k \right)}}$, $k=k+1$ ;
		\ENDIF
		\ENDWHILE
	\end{algorithmic}
\end{algorithm}

To simplify the notation, the augmented Lagrange function can be rewritten as
\begin{equation}
	\phi \left( {\boldsymbol{P},\boldsymbol{\lambda} ,\sigma } \right) = E\left( {\boldsymbol{P}} \right) - {{\boldsymbol{\lambda}}^{\rm T}}{\boldsymbol{h}}\left( {\boldsymbol{P}} \right) + \sigma {\boldsymbol{h}}{\left( {\boldsymbol{P}} \right)^{\rm T}}{\boldsymbol{h}}\left( {\boldsymbol{P}} \right),
\end{equation}
where ${\boldsymbol{\lambda }} = {\left[ {{\lambda _1},{\lambda _2}, \cdots ,{\lambda _{2M + N-1}}} \right]^{\rm T}}$, ${\boldsymbol{h}}\left( {\boldsymbol{P}} \right) = {\left[ {{h_1}\left( {\boldsymbol{P}} \right),{h_2}\left( {\boldsymbol{P}} \right), \cdots ,{h_{2M + N - 1}}\left( {\boldsymbol{P}} \right)} \right]^{\rm T}}$, ${h_1}\left( \boldsymbol{P} \right) = D - {D_{\min }}$, ${h_{j+1}}\left( \boldsymbol{P} \right) = \sum\nolimits_{i = 1}^M {{P_{i,j}}}  - {P_T},j = 1, 2, \ldots ,2M + N -2 $. The difference between $\phi \left( {\boldsymbol{P},\boldsymbol{\lambda} ,\sigma } \right)$ and Lagrange function $\mathcal{L}{\left( {{\boldsymbol{P}};{\boldsymbol{\lambda }}} \right)}$ is that $\phi \left( {\boldsymbol{P},\boldsymbol{\lambda} ,\sigma } \right)$ adds a penalty term $\sigma {\boldsymbol{h}}{\left( {\boldsymbol{P}} \right)^{\rm T}}{\boldsymbol{h}}\left( {\boldsymbol{P}} \right)$. This distinction makes the augmented Lagrange function have different properties from the Lagrange function. In contrast to the penalty function method, 
the penalty factor $\sigma$ of multiplier punitive method does not have to tend to infinity, as long as a sufficiently large $\sigma$ is taken, the optimal solution of the problem~\eqref{eq:36} can be obtained by minimizing $\phi \left( {\boldsymbol{P},\boldsymbol{\lambda} ,\sigma } \right)$.

Suppose that ${{\boldsymbol{\bar P}}}$ is a minimizer of $\phi \left( {{\boldsymbol{P}},{\boldsymbol{\bar \lambda }},\sigma } \right)$, where ${{\boldsymbol{\bar \lambda }}}$ is the optimal Lagrange multiplier. Then by the sufficient condition for the optimization problem, we have that $\nabla {\phi _{\boldsymbol{P}}}\left( {{\boldsymbol{\bar P}},{\boldsymbol{\bar \lambda }},\sigma } \right) = 0$, and for any direction ${\boldsymbol{d}} \in {\mathbb{R}^n}$, ${{\boldsymbol{d}}^T}\nabla \phi _{\boldsymbol{P}}^2\left( {{\boldsymbol{\bar P}},{\boldsymbol{\bar \lambda }},\sigma } \right){\boldsymbol{d}} > 0$. Since
\begin{equation}
\begin{aligned}
{\nabla _{\boldsymbol{P}}}\phi \left( {{\boldsymbol{\bar P}},{\boldsymbol{\bar \lambda }},\sigma } \right) & = \nabla E\left( {{\boldsymbol{\bar P}}} \right) - \sum\limits_{j = 1}^{2M + N - 1} {{{{\bar \lambda }}}_j}\nabla {h_j}\left( {{\boldsymbol{\bar P}}} \right) \\
&+ 2\sigma \sum\limits_{j = 1}^{2M + N - 1} {{h_j}\left( {\boldsymbol{P}} \right)\nabla {h_j}\left( {\boldsymbol{P}} \right)}   = 0,
\end{aligned}
\end{equation}
combined with ${h_j}\left( {{\boldsymbol{\bar P}}} \right) = 0,j = 1,2,  \cdots ,2M + N - 1$, yields
\begin{equation}
\nabla E\left( {{\boldsymbol{\bar P}}} \right) - 2\sum\limits_{j = 1}^{2M + N - 1} {{{{{\bar \lambda }}}_j}\nabla {h_j}\left( {{\boldsymbol{\bar P}}} \right)}  = 0.\label{sta}
\end{equation}
Thus, ${{\boldsymbol{\bar P}}}$ is a KKT point of the problem~\eqref{eq:36}.

By ${{\boldsymbol{d}}^T}\nabla \phi _{\boldsymbol{P}}^2\left( {{\boldsymbol{\bar P}},{\boldsymbol{\bar \lambda }},\sigma } \right){\boldsymbol{d}} > 0$, for any ${\boldsymbol{d}} \in {\mathbb{R}^n}$ such that ${{\boldsymbol{d}}^T}\nabla {h_j}\left( {{\boldsymbol{\bar P}}} \right) = 0,j = 1, 2, \cdots ,2M + N - 1$, we have
\begin{equation}
{{\boldsymbol{d}}^T}\left( {{\nabla ^2}E\left( {{\boldsymbol{\bar P}}} \right) - 2\sum\limits_{j = 1}^{2M + N - 1} {{{\bar \lambda }_j}{\nabla ^2}{h_j}\left( {{\boldsymbol{\bar P}}} \right)} } \right){\boldsymbol{d}} > 0.
\end{equation}
Therefore, ${{\boldsymbol{\bar P}}}$ is a local minimum of the original problem \eqref{E}.

Unfortunately, although the penalty factor can be sufficiently large, the value of the optimal Lagrange multiplier ${{\boldsymbol{\bar \lambda }}}$ cannot be known until the optimal solution is obtained. Therefore, it is necessary to investigate how to determine ${{\boldsymbol{\bar \lambda }}}$ and $\sigma$. Typically, $\boldsymbol{\lambda}$ is corrected during the iterative process based on a sufficiently large $\sigma$ and an initial estimate of the Lagrange multiplier $\boldsymbol{\lambda}$, trying to make $\boldsymbol{\lambda}$ converge to ${{\boldsymbol{\bar \lambda }}}$.

Assume that in the $k$th iteration, the optimal solution of $\phi \left( {\boldsymbol{P},\boldsymbol{\lambda^{{\left( k \right)}}} ,\sigma } \right)$ is attained at ${{\boldsymbol{P}}^{\left( k \right)}}$ , the estimate of $\boldsymbol{\lambda}$ in this iteration is ${{\boldsymbol{\lambda }}^{\left( k \right)}}$, and the penalty factor is $\sigma$, then we have ${\nabla _{\boldsymbol{P}}}\phi \left( {\boldsymbol{P^{{\left( k \right)}}},\boldsymbol{\lambda^{{\left( k \right)}}} ,\sigma } \right) = 0$.

For the optimal solution ${{\boldsymbol{\bar P}}}$ of the problem~\eqref{eq:36}, when $\nabla {h_1}\left( {{\boldsymbol{\bar P}}} \right)$, $\nabla {h_2}\left( {{\boldsymbol{\bar P}}} \right)$, $ \ldots $, $\nabla {h_{2M+N-1}}\left( {{\boldsymbol{\bar P}}} \right)$ are linearly independent, ${{\boldsymbol{\bar P }}}$ satisfying \eqref{sta}.

If ${{\boldsymbol{P}}^{\left( k \right)}} = {\boldsymbol{\bar P}}$, then ${{\bar \lambda }_j} = \lambda _j^{\left( k \right)} - 2\sigma {h_j}\left( {{{\boldsymbol{P}}^{\left( k \right)}}} \right)$. In general, ${{\boldsymbol{P}}^{\left( k \right)}} \ne {\boldsymbol{\bar P}}$ and \eqref{sta} does not hold, but $\boldsymbol{\lambda}$ can be corrected by
\begin{equation}
\lambda _j^{\left( {k + 1} \right)} = \lambda _j^{\left( k \right)} - 2\sigma {h_j}\left( {{{\boldsymbol{P}}^{\left( k \right)}}} \right),j = 1,2, \ldots ,2M + N - 1,
\end{equation}
then perform the $(k+1)$th iteration to find the minimizer of $\phi \left( {\boldsymbol{P},\boldsymbol{\lambda^{{\left( k+1 \right)}}} ,\sigma } \right)$. Repeat the process so that ${{\boldsymbol{\lambda }}^{\left( k \right)}} \to {\boldsymbol{\bar \lambda }} $, and ${{\boldsymbol{P}}^{\left( k \right)}} \to {\boldsymbol{\bar P}}$.

In this method, the penalty term is a penalty for the iteration point leaving the feasible region, which will force the iteration point to approach the feasible region during the minimization process. With the increase of the penalty factor, the distance between the iteration point and the optimal solution is closer. By adjusting the penalty factors continuously, the optimal solution can be found. The algorithm corrects the penalty factors and Lagrange multipliers based on the following considerations.

If ${\left\| {{\bf{h}}\left( {{{\bf{P}}^{\left( k \right)}}} \right)} \right\|_\infty } \ge {\left\| {{\bf{h}}\left( {{{\bf{P}}^{\left( {k - 1} \right)}}} \right)} \right\|_\infty }$, the iteration point tends to be far away from the constrained surface, thus the penalty term should be increased in the next iteration.

If ${\left\| {{\bf{h}}\left( {{{\bf{P}}^{\left( {k - 1} \right)}}} \right)} \right\|_\infty } > {\left\| {{\bf{h}}\left( {{{\bf{P}}^{\left( k \right)}}} \right)} \right\|_\infty } \ge \frac{1}{4}{\left\| {{\bf{h}}\left( {{{\bf{P}}^{\left( {k - 1} \right)}}} \right)} \right\|_\infty }$ and ${\sigma ^{\left( k \right)}} = {\sigma ^{\left( {k - 1} \right)}}$, the proximity of the iterative point to the constrained surface is not significant, which is related to the unadjusted penalty factor.

If ${\left\| {{\bf{h}}\left( {{{\bf{P}}^{\left( k \right)}}} \right)} \right\|_\infty } < \frac{1}{4}{\left\| {{\bf{h}}\left( {{{\bf{P}}^{\left( {k - 1} \right)}}} \right)} \right\|_\infty }$, the iterative point approaches the constrained surface significantly. In the next iteration, only the Lagrange multipliers need to be adjusted.

If ${\left\| {{\bf{h}}\left( {{{\bf{P}}^{\left( {k - 1} \right)}}} \right)} \right\|_\infty } > {\left\| {{\bf{h}}\left( {{{\bf{P}}^{\left( k \right)}}} \right)} \right\|_\infty } \ge \frac{1}{4}{\left\| {{\bf{h}}\left( {{{\bf{P}}^{\left( {k - 1} \right)}}} \right)} \right\|_\infty }$ and ${\sigma ^{\left( k \right)}} > {\sigma ^{\left( {k - 1} \right)}}$, it means that in this iteration, increasing the penalty factor has an effect on the proximity of the current iteration point to the constrained surface, but the effect is not significant. In the next iteration, only the Lagrange multipliers are adjusted.

The proposed multiplier punitive function-based algorithm is presented in Algorithm~\ref{alg:1}, which can compute the approximate optimal solution to problem~\eqref{E}.

\section{Performance Evaluation}\label{S4}

In this section, the performance of the proposed energy efficiency scheme based on multiplier punitive function algorithm is evaluated with various critical parameters, including the number of MRs, the inter-RRH distance, and the velocity of the train. The train velocity estimation error, which is assumed to be zero in Section~\ref{S2-1}, is also taken into account in the simulations. Specifically, we make a comparison of the proposed scheme with four baseline schemes to demonstrate the optimization gain.

\subsection{Simulation Setup}\label{S4-1}

\begin{figure*}[!t]
  \centering 
  \includegraphics[width=6.4in]{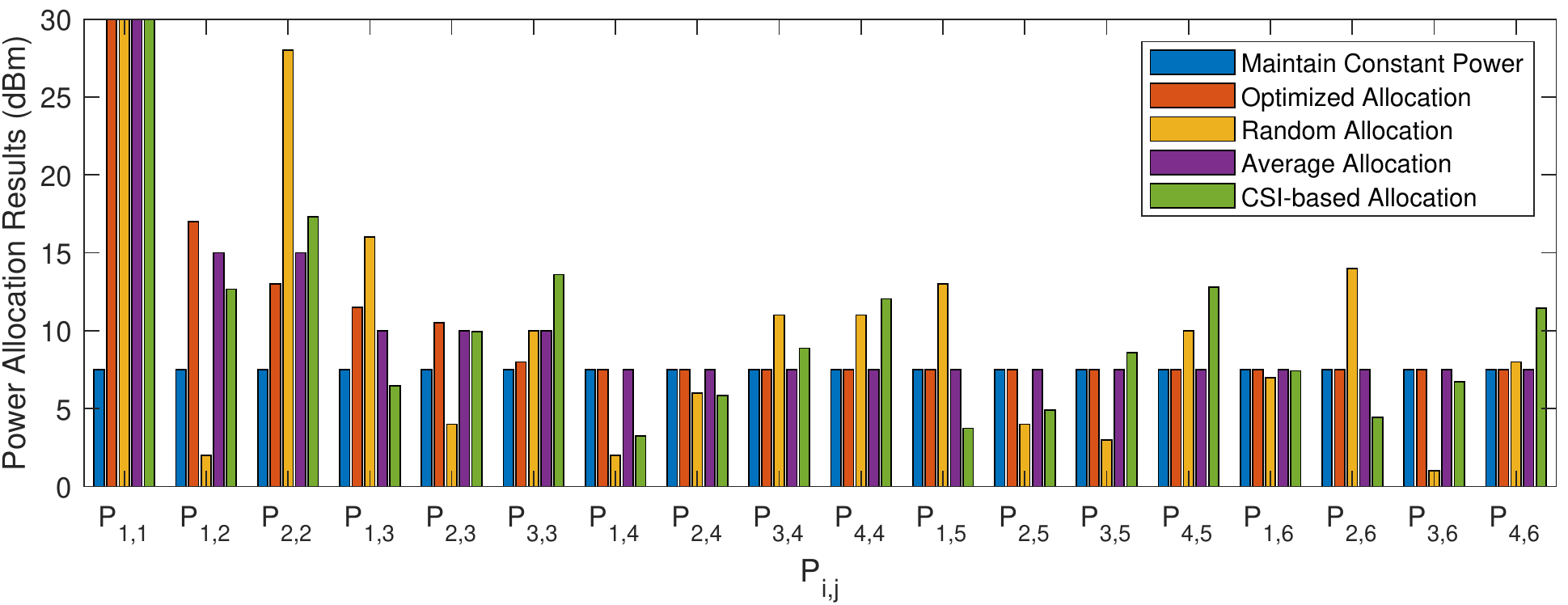}
  \caption{Power allocation results at each segment.}
  \label{allocation}
 \end{figure*}

\linespread{1.2}
\begin{table}[!t]
\begin{center}
\caption{Simulation Parameters}
\begin{tabular}{lcc}
\toprule
{\em Parameter} & {\em Symbol}  & {\em Value} \\
\midrule
Distance from a RRH to the rail & $d_0$ & 20 m  \\
Angle of the half-power beamwidth & $\theta_{-3dB}$ & $30^{\circ}$ \\
Shadowing margin &  $\xi$  & 10 dB \\
Path loss exponent & $n$ & 2 \\
Carrier wavelength & $\lambda$ & 5 mm \\
System bandwidth&  $B$ & 2.16 GHz \\
Noise figure  &  $NF$   &  6 dB \\
Transmit power budget &  $P_T$  & 30 dBm, 40dBm \\
Distance between MRs  &   $d_{MR}$ &    25 m \\
Number of location bins in the second stage  &   $N$ &    6\\
\bottomrule
\end{tabular}
\label{table1}
\end{center}
\end{table}

The realistic directional antenna model supported in standards like IEEE 802.15.3c is adopted here similar to \cite{lei}, where the linear scale of the main lobe is Gaussian and sidelobes are constant \cite{35}. The directional antenna gain $G(\theta)$ in $dB$ is represented as
\begin{equation}
G(\theta)=\left\{
\begin{array}{ll}
G_0-3.01 \cdot {\left(\dfrac{2\theta}{\theta_{-3dB}}\right)}^{2},& {0^\circ\leq\theta\leq{\theta_{ml}}/2}, \\ 
G_{sl},& {{\theta_{ml}}/2\leq\theta\leq180^\circ},
\end{array}
\right.\label{eq2}
\end{equation}
where $\theta$ denotes an arbitrary angle depends on the actual direction between the transceivers, $\theta_{-3dB}$ is the half-power beamwidth angle, and $\theta_{ml}$ is the main lobe beamwidth, which satisfies $\theta_{ml}=2.6\cdot\theta_{-3dB}$.
In~\eqref{eq2}, $G_0$ denotes the maximum antenna gain, given by
\begin{equation}
G_0=10\log{\left(\dfrac{1.6162}{\sin{\left(\theta_{-3dB}/2\right)}}\right)^2}.\label{eq3}
\end{equation}
while the sidelobe gain $G_{sl}$ is written as
\begin{equation}
G_{sl}=-0.4111\cdot{\ln(\theta_{-3dB})-10.579}.\label{eq4}
\end{equation}

Consider the network model shown in Fig. \ref{network}, and main simulation parameters are given in Table I. 

\subsection{Comparison with Baseline Schemes}\label{S4-2}

To evaluate performance of the proposed scheme and further illustrate the optimization gain, we choose the following four baseline schemes for comparison purpose:

\begin{itemize}
\item {\em Maintain constant transmit power}: once the MR enters the coverage area of the cell, a constant transmission power of
${P_T}/M$ is allocated to it, and the power is maintained until the MR leaves the cell. 

\item {\em Random transmit power allocation}: since the optimal solution is obtained when the constraints are equal as in~\eqref{eq:36}, it is reasonable to allocate transmit power randomly when the constraints~\eqref{36-2} are satisfied.

\item {\em Average transmit power allocation}: when the train enters or leaves the cell, $P_T$ is divided and allocated equally to the MRs in the cell, and when MRs are all traveling in the cell, $P_T$ is allocated equally to the $M$ MRs.

\item {\em Channel state information (CSI)-based power allocation}: the transmit power is allocated according to the coefficient determined by the CSI. Specifically, the transmit power allocated to the $i$th MR during $[t_{j-1},t_j)$ is given by ${P_{i.j}} = {{{{\left( {{{\left| {{h_{i,j}}} \right|}^2}} \right)}^{ -\alpha}}} \mathord{\left/{\vphantom {{{{\left( {{{\left| {{h_{i,j}}} \right|}^2}} \right)}^{ - \alpha}}} {\sum\nolimits_{i = 1}^M {{{\left( {{{\left| {{h_{i,j}}} \right|}^2}} \right)}^{ - 1}}} }}} \right.\kern-\nulldelimiterspace} {\sum\nolimits_{i = 1}^M {{{\left( {{{\left| {{h_{i,j}}} \right|}^2}} \right)}^{ - \alpha}}} }} {P_T},\forall j = {1,2, \cdots ,2M + N - 1}$, where $\alpha$ is a constant coefficient and is set to $0.2$.
\end{itemize}

The {\em Optimized transmit power allocation} refers to transmit power allocation through the proposed multiplier punitive function-based algorithm.

Fig. \ref{allocation} plots the power allocation results at each segment with $d_l=200m$, $v=300km/h$, $P_T=30dBm$ and $M=4$, where $P_{i,j}\ (j=1,2,3)$ denotes the allocated power of the train entering the cell, and $P_{i,j}\ (j=4,5,6)$ denotes the allocated power when all MRs are covered in the cell. Only half of the results are shown in the figure considering the symmetry and only a random allocation result is provided. Comparing the optimized scheme and the average allocation scheme, the power allocation of the two is similar after the MRs are all covered in the cell. On the one hand, from the time MR enters the location bin to the time it leaves the location bin, the short length of each location bin results in a small change in the distance between the MR and the RRH. Based on the adopted channel model, the MR experiences similar channel conditions in different location bins and is allocated the same power level. On the other hand, the distance between adjacent MRs is also a small value. Similarly, multiple MRs with nearly similar channel gains are allocated equal levels of power. In particular, when the train gradually enters the cell, the first MR of the optimized scheme is usually allocated more power, since the first MR is closer to the RRH, more power allocated to the first MR helps transmit more data. It can also be found from Fig. \ref{allocation} that for the CSI-based scheme, MRs with low channel gains are allocated to high transmit power levels, while MRs with high channel gains are allocated to low power levels.

\begin{figure}[t!]
    \begin{center}
        \includegraphics[width=2.7in]{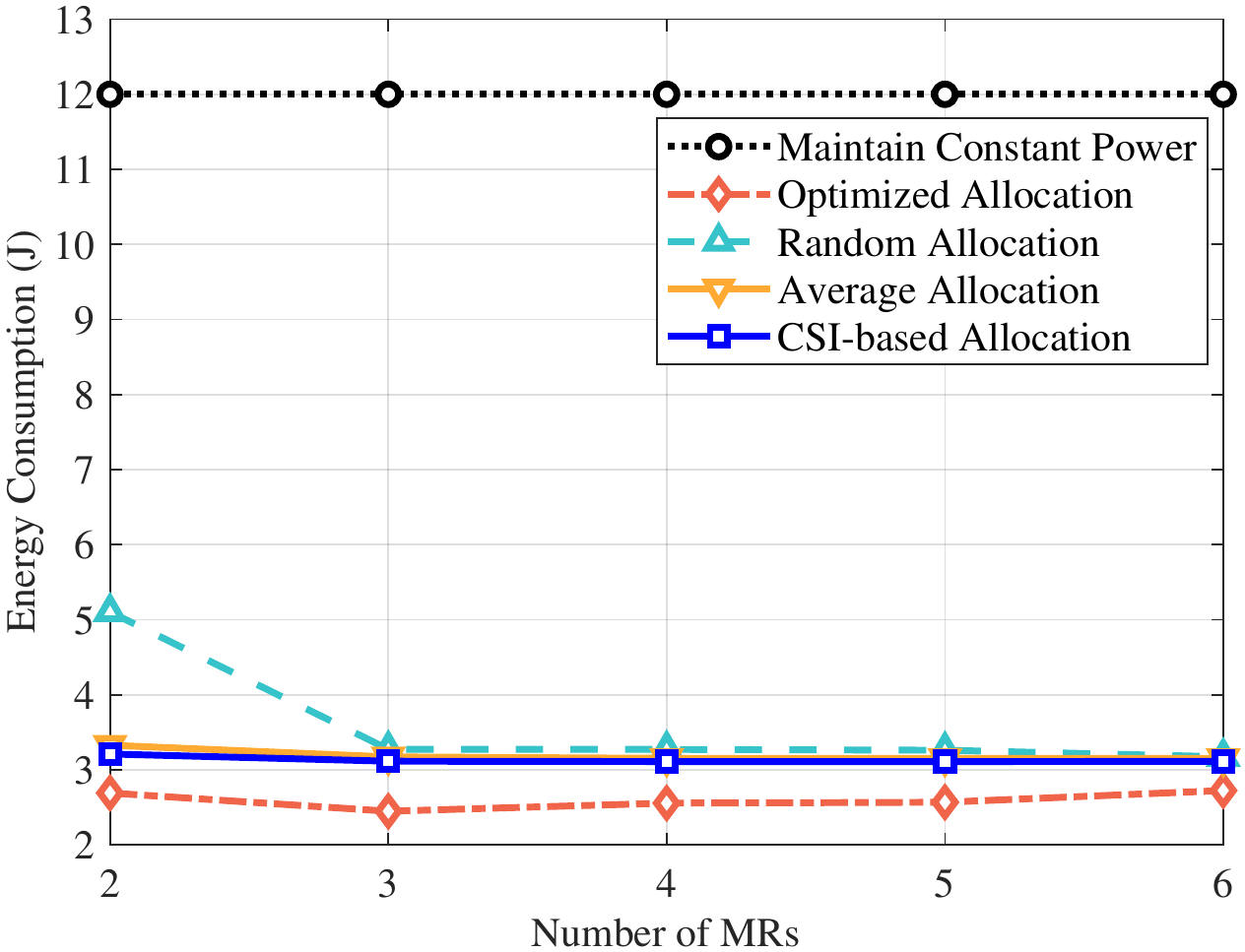}
        \caption{Energy consumption comparison under different $M$ values.}
        \label{EM40}
    \end{center}
\end{figure}

Fig. \ref{EM40} plots the energy consumption comparison of the five schemes under different $M$ values with $d_l=200m$, $v=300km/h$ and $P_T=40dBm$. From the figure, it is observed that when the $M$ increases, the energy consumption of the optimized scheme remains at a low level, and the average allocation scheme and the CSI-based allocation scheme exhibit similar performance. Compared with the average and CSI-based allocation scheme, the optimized scheme has better performance in reducing energy consumption. Due to the random power allocation, its energy consumption exhibits slight fluctuates. However, overall it is lower than the energy consumption value of the constant power scheme. Therefore, when the number of MRs is changed, the optimized, random, average, and CSI-based allocation schemes all reduce the energy consumption of train-ground communications, while the proposed scheme can achieve better performance.

\begin{figure}[t!]
    \begin{center}
        \includegraphics[width=2.7in]{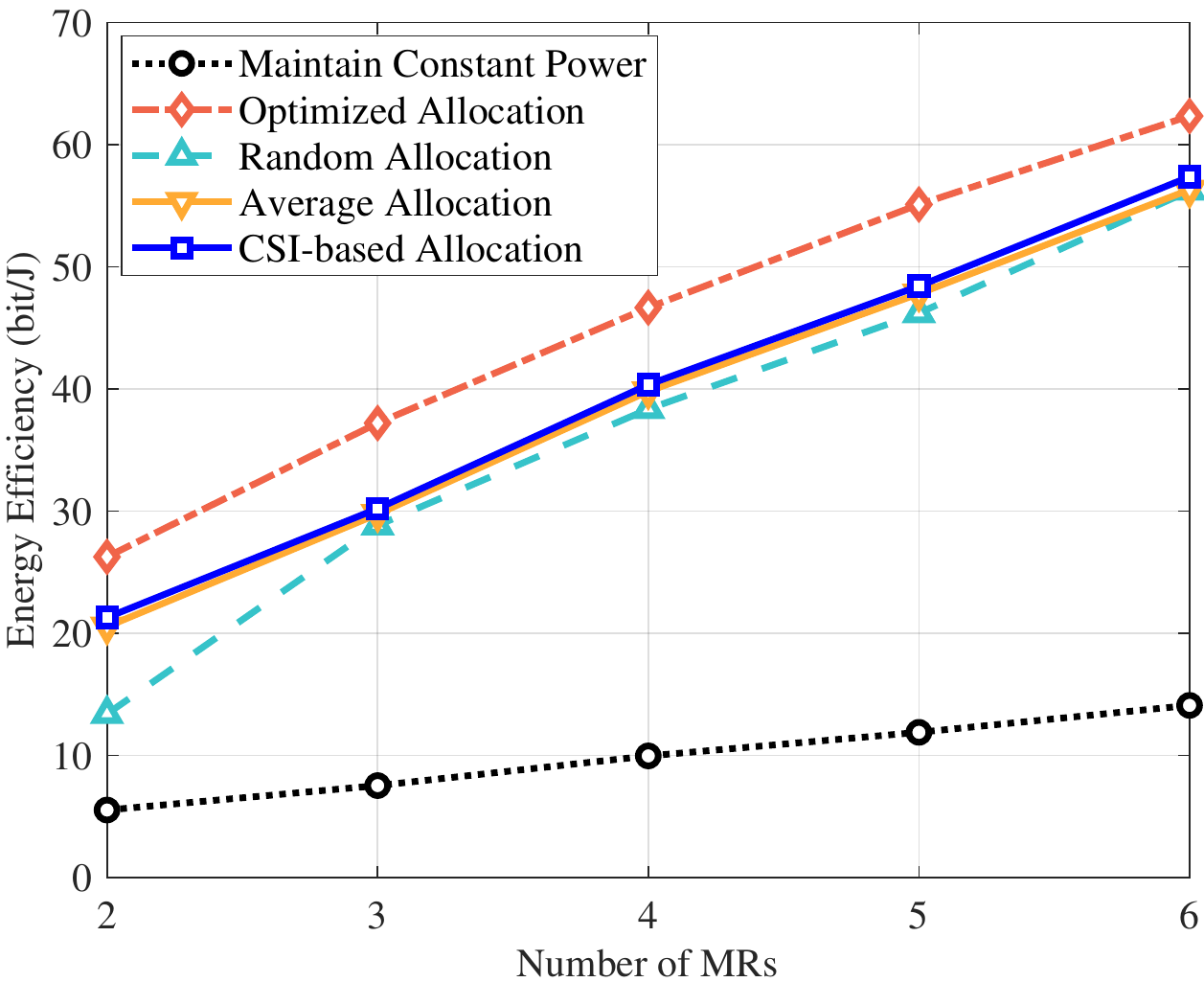}
        \caption{Energy efficiency comparison under different $M$ values.}
        \label{EEM40}
    \end{center}
\end{figure}

The results of energy efficiency varying with the number of MRs are shown in Fig.~\ref{EEM40}. By comparison, we can find that more installed MRs do help improve the energy efficiency of train-ground mmWave communications. The energy efficiencies of the constant, optimized, average and CSI-based allocation schemes are all improved as more MRs are used. The value of the random allocation scheme exhibits a generally increasing trend, although with fluctuations. It is worth noting that the proposed scheme achieves the highest energy efficiency, and the gap with the average allocation scheme and the CSI-based scheme is obvious. In particular, the energy efficiencies of random power allocation are similar to the values of the average allocation scheme, the random allocation scheme can also achieve good performance in some cases.

\begin{figure}[t]
   \begin{center}
        \includegraphics[width=2.7in]{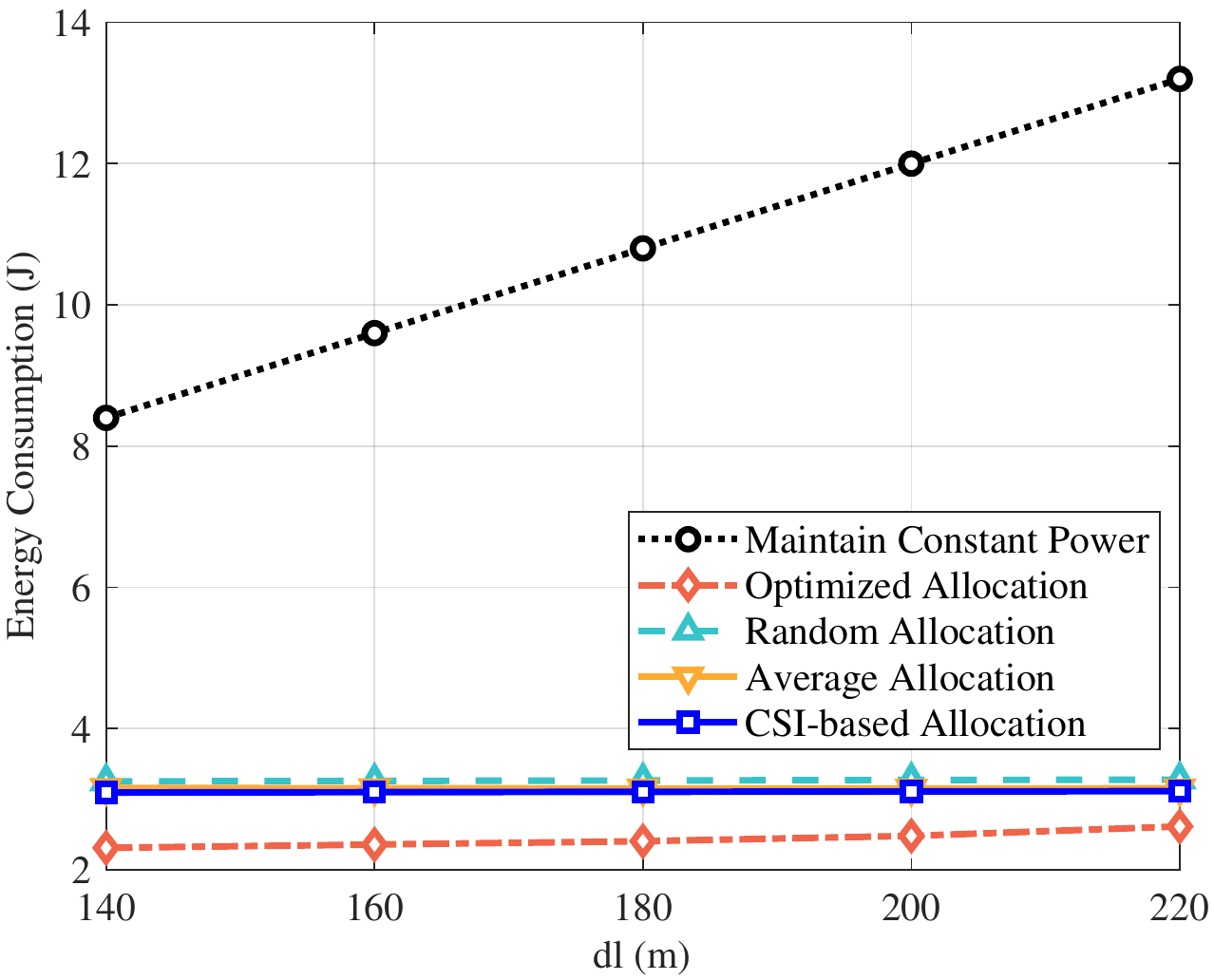}
        \caption{Energy consumption comparison under different $d_l$ values.}
        \label{Edl}
    \end{center}
\end{figure}

\begin{figure}[t!]
    \begin{center}
        \includegraphics[width=2.7in]{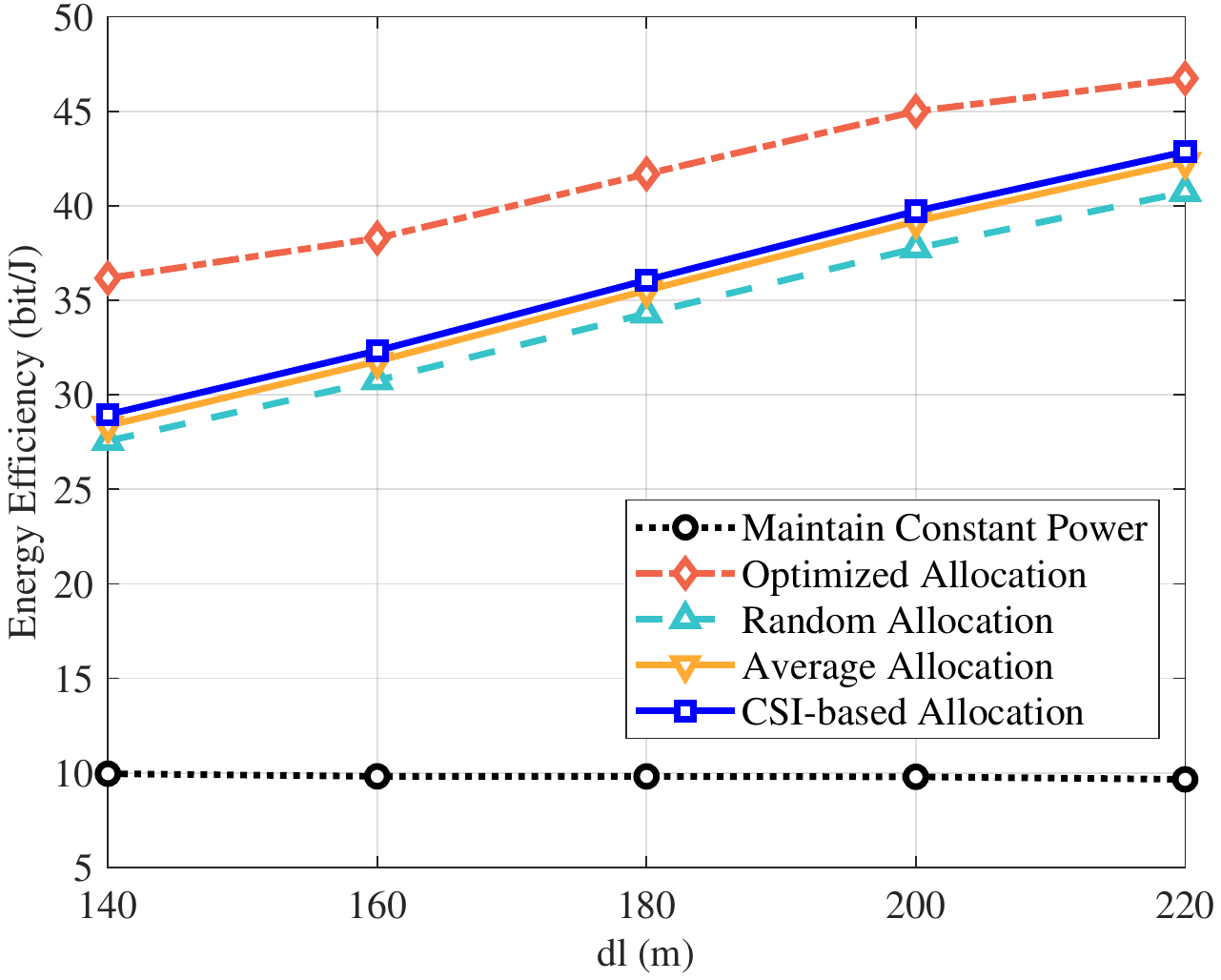}
        \caption{Energy efficiency comparison under different $d_l$ values.}
        \label{EEdl}
    \end{center}
\end{figure}

Fig.~\ref{Edl} shows the energy consumption performance of the five schemes under different $d_l$ values. We set the transmit power as $40\ dBm$, the speed of the train as $300km/h$, and the number of MRs as 4. It is observed that the energy consumption of the constant power scheme rises quickly at a constant rate with increased $d_l$. In contrast, the energy consumption values of the other four schemes don't change much with the increased $d_l$. The proposed scheme outperforms the other four schemes in general. To be specific, the optimized scheme saves about $79.6\%$ energy compared with the constant power scheme when $d_l=200m$. The energy consumption of the optimized scheme is also lower than that of the average allocation scheme, the CSI-based allocation scheme, and the random allocation scheme for the entire range of $d_l$ values examined.

Energy efficiencies of the five schemes under different $d_l$ values are shown in Fig.~\ref{EEdl}. The energy efficiency of the constant power scheme decreases with the increase of $d_l$, while the energy efficiencies of the other four schemes, on the contrary, all grow as $d_l$ is increased. Similarly, the proposed scheme outperforms all the four baseline schemes for the entire range of $d_l$ values examined. In particular, compared with the CSI-based scheme, the proposed scheme improves the energy efficiency by $25.54\%$ when $d_l=140m$. The CSI-based scheme performs slightly better compared to the average scheme.

\begin{figure}[t!]
    \begin{center}
        \includegraphics[width=2.7in]{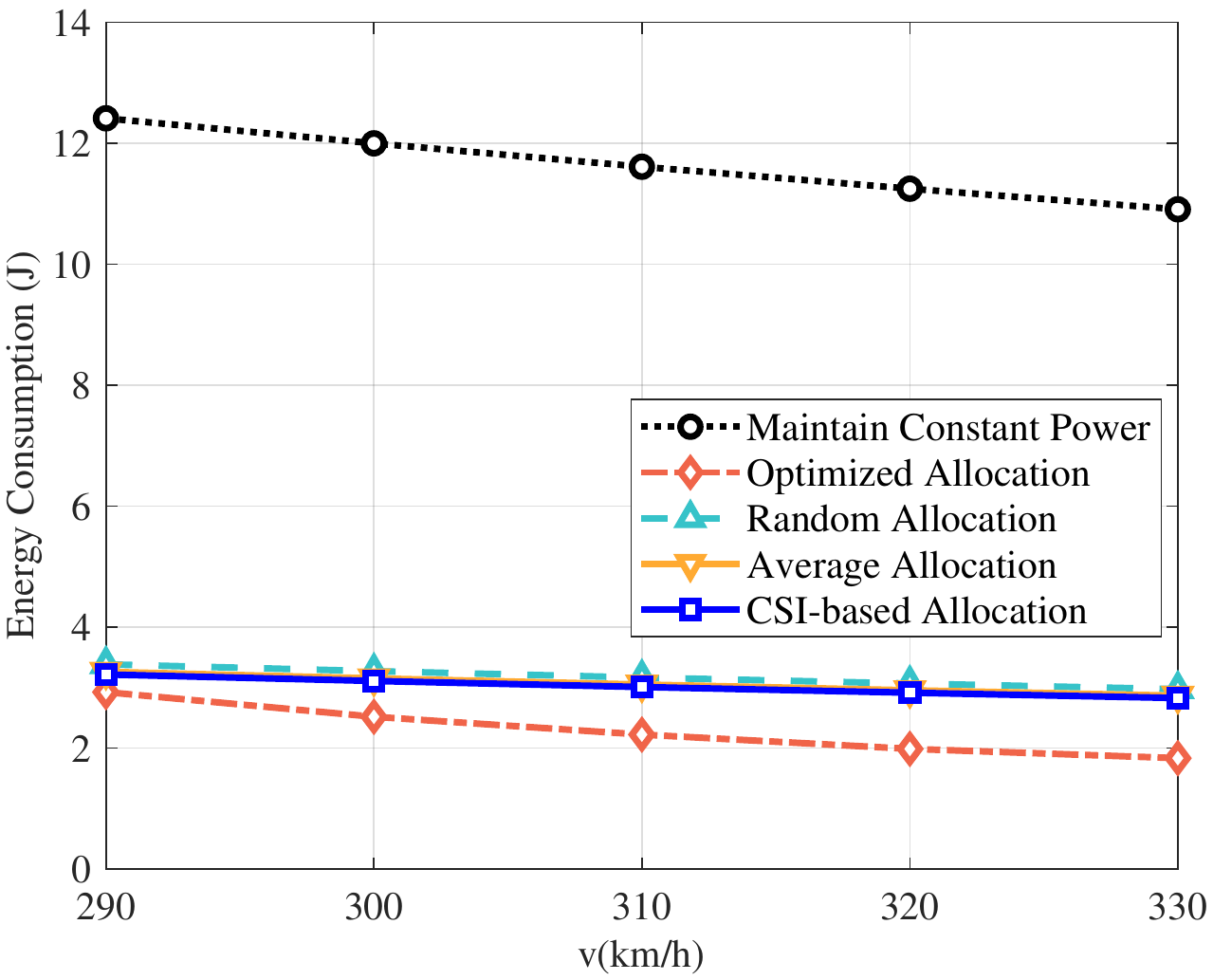}
        \caption{Energy consumption comparison under different $v$ values.}
        \label{EV30}
    \end{center}
\end{figure}

The performance of the five schemes in terms of velocity with $M=4$, $d_l=200m$, and $P_T=40dBm$ is shown in Fig.~\ref{EV30}. From the figure, the optimization scheme has lower energy consumption compared with the other four baseline schemes. The energy consumptions of five schemes all decrease with the increase of train speed. As the train velocity rises, the time it takes for the train to travel the same distance is reduced, and thus for the specific transmit power allocation method, the overall energy consumption is reduced. In particular, the energy consumption of the proposed optimization scheme is always lower than the value of the random allocation scheme, the average allocation scheme, and the CSI-based allocation scheme, and the optimization gain becomes more apparent as the train speed goes up.

\begin{figure}[t!]
    \begin{center}
        \includegraphics[width=2.7in]{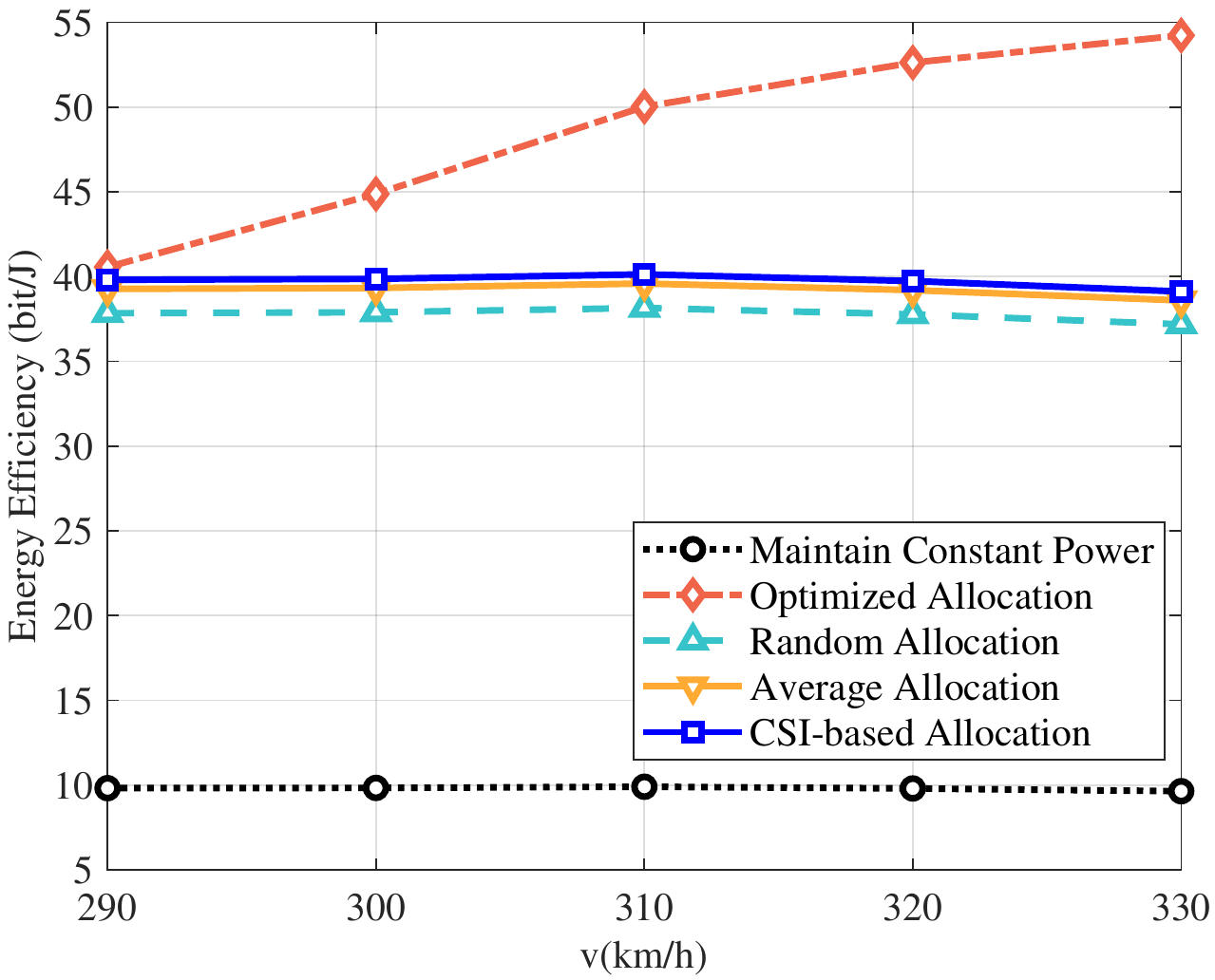}
        \caption{Energy efficiency comparison under different $v$ values.}
        \label{EEV30}
    \end{center}
\end{figure}

The energy efficiencies with varying velocities is shown in Fig.~\ref{EEV30}. When the train speed is increased, the energy efficiencies of the optimized scheme, the random allocation scheme, the average allocation scheme, and the CSI-based allocation scheme all remain a high level. The energy efficiency of the proposed optimization scheme grows gradually and the value is higher than the other four baseline schemes. The difference between the optimization scheme and the CSI-based scheme is more apparent when $v>290km/h$. Thus as the train speed rises, the energy efficiency optimization effect of the optimization scheme becomes better.

\begin{figure}[t]
    \begin{center}
        \includegraphics[width=2.7in]{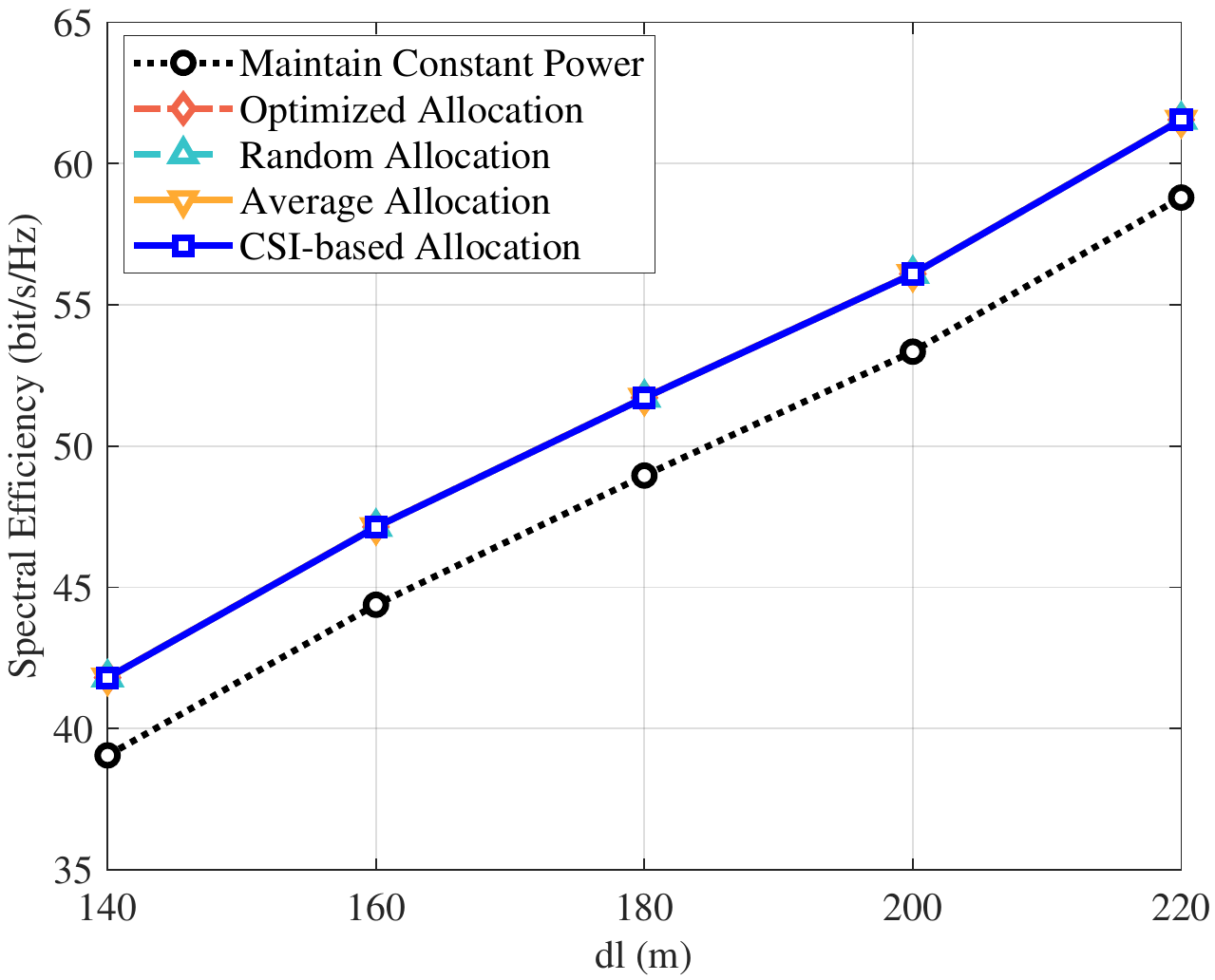}
        \caption{Spectral efficiency comparison under different $d_l$ values.}
        \label{throughtpd}
    \end{center}
\end{figure}

\begin{figure}[!t]
    \begin{center}
        \includegraphics[width=2.7in]{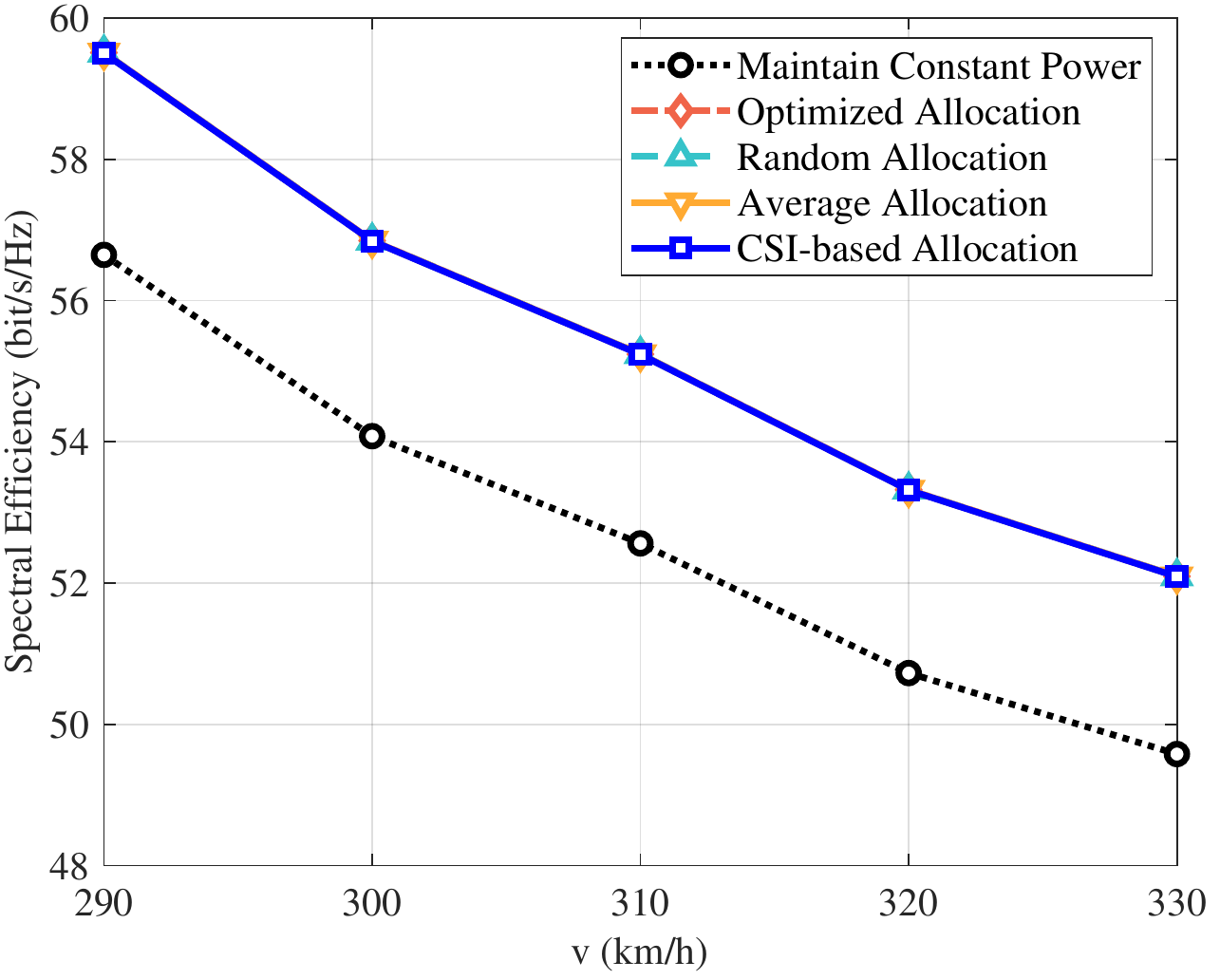}
        \caption{Spectral efficiency comparison under different $v$ values.}
        \label{throughtpv}
    \end{center}
\end{figure}
In addition to energy consumption and energy efficiency, we also compare the spectral efficiency of the five schemes. Fig.~\ref{throughtpd} provides the spectral efficiency performance curves as $d_l$ increases and Fig.~\ref{throughtpv} shows the impact of velocity on spectral efficiency. The optimized, random, average, and CSI-based schemes have similar performance, and they all outperform the constant scheme. In Fig.~\ref{Edl}, Fig.~\ref{EEdl}, Fig.~\ref{EV30}, and Fig.~\ref{EEV30}, the proposed scheme achieves the highest energy efficiency and the lowest energy consumption, while in Fig.~\ref{throughtpd} and Fig.~\ref{throughtpv}, the proposed scheme can only reach similar levels of spectral efficiency as the average scheme, the random scheme, and the CSI-based scheme. The results are related to the optimization model of this paper. Problem \eqref{E} optimizes the energy efficiency by minimizing the energy consumption under the constraints of the transmitted data volume and the transmit power. On the one hand, the proposed scheme achieves better $EE$ for a given transmitted data volume $D_{min}$ by reducing the transmit power. On the other hand, precisely because of the strict constraints on the total transmitted data and transmit power, the system cannot achieve a higher spectral efficiency. Therefore, there is a trade-off between the energy efficiency and spectral efficiency of the system, which suggests that the optimal transmit power should be allocated with a full consideration of energy efficiency and spectral efficiency.

\subsection{Computational Complexity Analysis}\label{S4-4}
Computational complexity is an important indicator to measure the performance of the algorithm. This section compares the computational complexity of the proposed scheme with the four benchmark schemes.

The constant scheme allocates the fixed transmit power of ${P_T}/M$ for the $i$th MR in the cell, the random scheme allocates a random power value between 0 and ${P_T}$ to each MR subject to the constraints \eqref{36-2}, and the average scheme allocates an equal power to each MR in the cell. All three schemes are based on straightforward power allocation criteria that do not depend on channel conditions, and are therefore physically easier to implement. The formulated problem is to allocate transmit power to each MR during each interval $\left[ {{t_{j - 1}},{t_j}} \right)$, containing a total of $M(M+N-1)$ variables, hence the computational complexity of all three schemes is $O\left( { M \left( {M + N - 1} \right)} \right)$.

The CSI-based scheme takes into account the channel fading. For a given $i$ and $j$, finding $P_{i,j}$ requires computing the coefficients ${{{{\left( {{{\left| {{h_{i,j}}} \right|}^2}} \right)}^{ -\alpha}}} \mathord{\left/{\vphantom {{{{\left( {{{\left| {{h_{i,j}}} \right|}^2}} \right)}^{ - \alpha}}} {\sum\nolimits_{i = 1}^M {{{\left( {{{\left| {{h_{i,j}}} \right|}^2}} \right)}^{ - \alpha}}} }}} \right.\kern-\nulldelimiterspace} {\sum\nolimits_{i = 1}^M {{{\left( {{{\left| {{h_{i,j}}} \right|}^2}} \right)}^{ - \alpha}}} }}$. The complexity of computing ${\sum\nolimits_{i = 1}^M {{{\left( {{{\left| {{h_{i,j}}} \right|}^2}} \right)}^{ - \alpha }}} }$ is $O\left(M\right)$, and therefore the complexity of computing $P_{i,j}$ is also $O\left(M\right)$. Considering all $M(M+N-1)$ variables, the complexity of the CSI-based scheme is $O\left( { M^2\left( {M + N - 1} \right)} \right)$.

The proposed algorithm first uses the gradient descent method to obtain the optimal solution of the augmented Lagrange function. The optimization problem \eqref{E} has $M(M+N-1)$ optimization variables. Denote the number of iterations of gradient descent method as $k_{GD}$, then the complexity of the gradient descent algorithm is $O\left( {{k_{GD}}  M  \left( {M + N - 1} \right)} \right)$. Let the number of iterations of the multiplier punitive function algorithm be $k_{MPF}$, thus the complexity of the proposed algorithm is $O\left( {{k_{MPF}}  {k_{GD}}  M  \left( {M + N - 1} \right)} \right)$.

In contrast, the constant, random, and average schemes have the lowest computational complexity but fail to achieve similar energy consumption and energy efficiency performance as the proposed scheme. The computational complexity of the proposed scheme is comparable to that of the CSI-based scheme, while the proposed scheme outperforms the CSI-based scheme in terms of energy consumption and energy efficiency performance.

\subsection{Impact of Velocity Estimation Error}\label{S4-3}

\begin{figure}[t!]
    \begin{center}
        \includegraphics[width=2.7in]{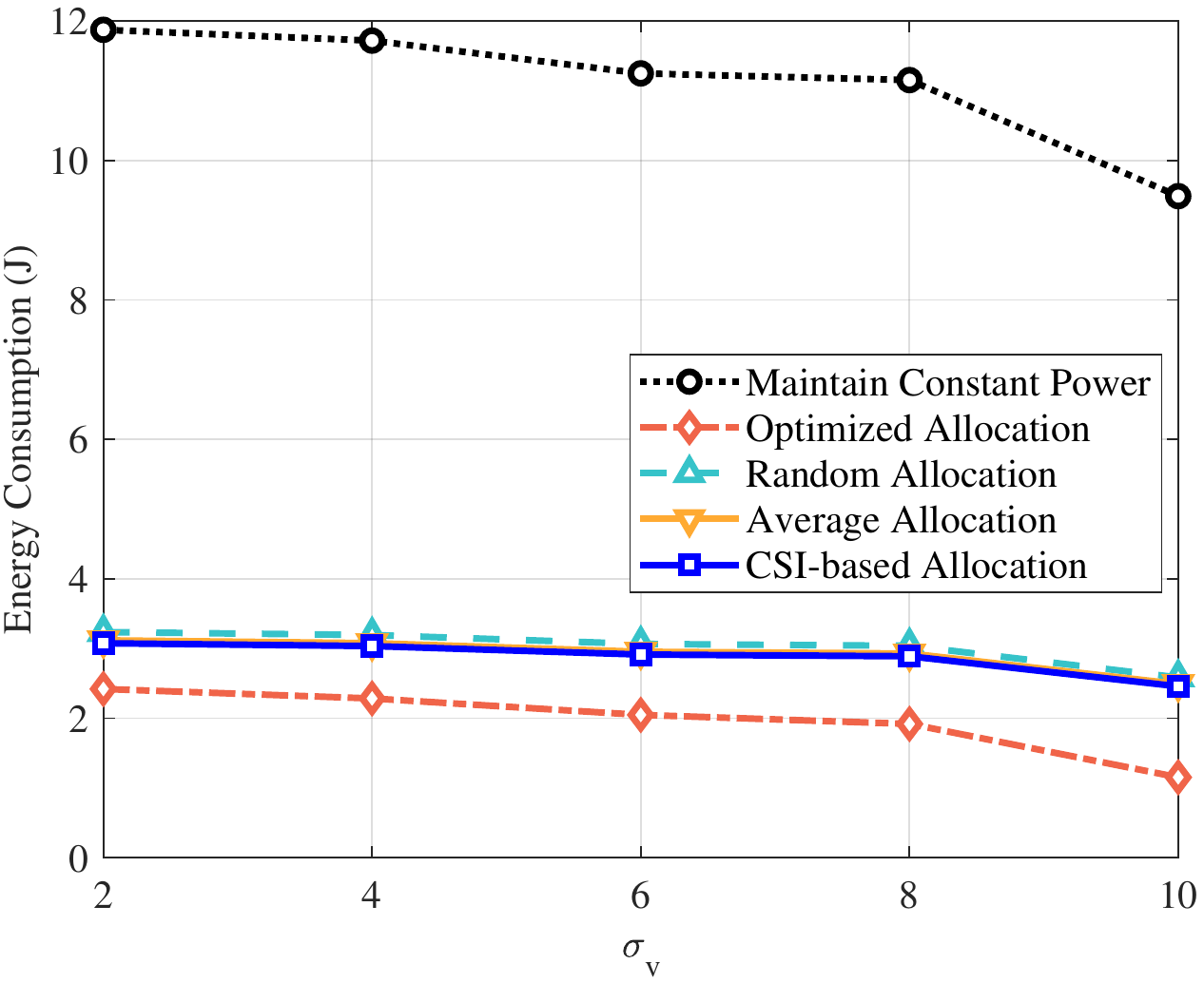}
        \caption{Energy consumption comparison under different $\sigma_v$ values.}
        \label{sigmaE}
    \end{center}
\end{figure}

\begin{figure}[t!]
    \begin{center}
        \includegraphics[width=2.7in]{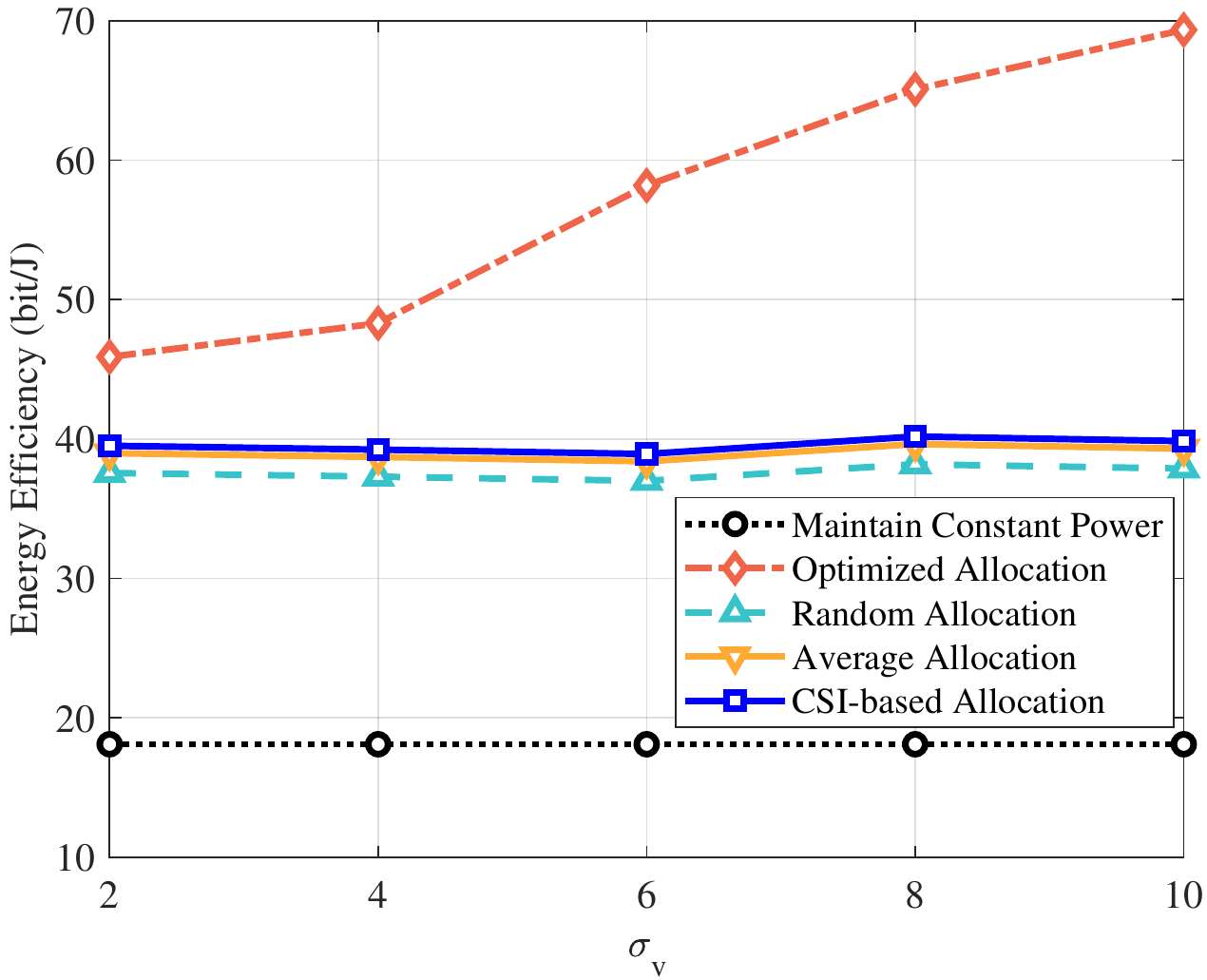}
        \caption{Energy efficiency comparison under different $\sigma_v$ values.}
        \label{sigmaEE}
    \end{center}
\end{figure}

Due to insufficient velocity information update as well as the additional latency for sending this information may cause velocity estimation error, now we take into account the velocity estimation error in our simulations. 
The velocity estimation error is assumed to follow Gaussian distribution here, i.e., $\hat{v}=v+v_e$, where $\hat{v}$ is the estimated velocity of the train, $v$ is the actual velocity and estimation error $v_e\sim\mathcal  N(0,\sigma_v^2)$ \cite{34}. $v_e$ are generated as positive numbers in the simulation to correspond to the results in Fig.~\ref{EV30} and Fig.~\ref{EEV30}.

Fig.~\ref{sigmaE} plots the energy consumptions of the five schemes under different $\sigma_v$ values. It can be obtained that when taking velocity estimation error into account, the energy consumption trends of the five schemes are the same as in Fig.~\ref{EV30}, and the proposed scheme achieves the lowest energy consumption. The optimized scheme is also the best choice.

Fig.~\ref{sigmaEE} plots the energy efficiency comparison of the five schemes under different $\sigma_v$ values. As we see, the energy efficiency of the proposed scheme continues to rise with increasing $\sigma_v$ values and achieves the highest energy efficiency as in Fig.~\ref{EEV30}. This shows that the velocity estimation error has little effect on the energy efficiency optimization of the proposed scheme, and the performance of the proposed scheme is still optimal under different errors. Therefore, the proposed scheme has a stable optimization effect.

\section{Conclusion}\label{S5}

This paper studies the energy efficiency problem of a mmWave communication network for HSRs with multiple MRs on top of the train. To achieve green train-ground communications, an energy consumption minimization problem is formulated subject to the total transmitted data and transmit power budget constraints. Aimed at this non-convex non-linear problem, a multiplier punitive function-based algorithm is proposed to allocated transmit power of MRs. Simulation results have demonstrated that the proposed scheme can achieve higher energy efficiency compared to four baseline schemes under various parameters, including the number of MRs, the inter-RRH distance and the speed of the train. We also show the good performance of the proposed algorithm when considering the speed estimation error.

\bibliographystyle{IEEEtran}
\bibliography{mybib}

\begin{thebibliography}{10}
\providecommand{\url}[1]{#1}
\csname url@samestyle\endcsname
\providecommand{\newblock}{\relax}
\providecommand{\bibinfo}[2]{#2}
\providecommand{\BIBentrySTDinterwordspacing}{\spaceskip=0pt\relax}
\providecommand{\BIBentryALTinterwordstretchfactor}{4}
\providecommand{\BIBentryALTinterwordspacing}{\spaceskip=\fontdimen2\font plus
\BIBentryALTinterwordstretchfactor\fontdimen3\font minus
  \fontdimen4\font\relax}
\providecommand{\BIBforeignlanguage}[2]{{%
\expandafter\ifx\csname l@#1\endcsname\relax
\typeout{** WARNING: IEEEtran.bst: No hyphenation pattern has been}%
\typeout{** loaded for the language `#1'. Using the pattern for}%
\typeout{** the default language instead.}%
\else
\language=\csname l@#1\endcsname
\fi
#2}}
\providecommand{\BIBdecl}{\relax}
\BIBdecl

\bibitem{3}
T.~{Zhou}, C.~{Tao}, and L.~{Liu}, ``{LTE}-assisted multi-link {MIMO} channel
  characterization for high-speed train communication systems,'' \emph{IEEE
  Transactions on Vehicular Technology}, vol.~68, no.~3, pp. 2044--2051, Mar.
  2019.

\bibitem{4}
Y.~{Dong}, C.~{Zhang}, P.~{Fan}, and P.~{Fan}, ``Power-space functions in high
  speed railway wireless communications,'' \emph{Journal of Communications and
  Networks}, vol.~17, no.~3, pp. 231--240, Jun. 2015.

\bibitem{1}
B.~{Ai}, A.~F. {Molisch}, M.~{Rupp}, and Z.~{Zhong}, ``{5G} key technologies
  for smart railways,'' \emph{Proceedings of the IEEE}, vol. 108, no.~6, pp.
  856--893, Jun. 2020.

\bibitem{7}
M.~{Xiao}, S.~{Mumtaz}, Y.~{Huang}, L.~{Dai}, Y.~{Li}, M.~{Matthaiou}, G.~K.
  {Karagiannidis}, E.~{Björnson}, K.~{Yang}, C.~{I}, and A.~{Ghosh},
  ``Millimeter wave communications for future mobile networks,'' \emph{IEEE
  Journal on Selected Areas in Communications}, vol.~35, no.~9, pp. 1909--1935,
  Sept. 2017.

\bibitem{12}
G.~{Yang} and M.~{Xiao}, ``Performance analysis of millimeter-wave relaying:
  Impacts of beamwidth and self-interference,'' \emph{IEEE Transactions on
  Communications}, vol.~66, no.~2, pp. 589--600, Feb. 2018.

\bibitem{17}
M.~{Giordani}, M.~{Mezzavilla}, and M.~{Zorzi}, ``Initial access in {5G}
  {mmWave} cellular networks,'' \emph{IEEE Communications Magazine}, vol.~54,
  no.~11, pp. 40--47, Nov. 2016.

\bibitem{16}
P.~{Raviteja}, Y.~{Hong}, and E.~{Viterbo}, ``Millimeter wave analog
  beamforming with low resolution phase shifters for multiuser uplink,''
  \emph{IEEE Transactions on Vehicular Technology}, vol.~67, no.~4, pp.
  3205--3215, Apr. 2018.

\bibitem{power-level}
S.~Hu, X.~Chen, W.~Ni, X.~Wang, and E.~Hossain, ``Modeling and analysis of
  energy harvesting and smart grid-powered wireless communication networks: A
  contemporary survey,'' \emph{IEEE Transactions on Green Communications and
  Networking}, vol.~4, no.~2, pp. 461--496, June 2020.

\bibitem{greenj}
A.~{Abrol} and R.~K. {Jha}, ``Power optimization in {5G} networks: A step
  towards green communication,'' \emph{IEEE Access Journal}, vol.~4, no.~1, pp.
  1355--1374, Apr. 2016.

\bibitem{ra1}
H.~Zhang, H.~Liu, J.~Cheng, and V.~C.~M. Leung, ``Downlink energy efficiency of
  power allocation and wireless backhaul bandwidth allocation in heterogeneous
  small cell networks,'' \emph{IEEE Transactions on Communications}, vol.~66,
  no.~4, pp. 1705--1716, Apr. 2018.

\bibitem{eh1}
S.~Zhang, N.~Zhang, S.~Zhou, J.~Gong, Z.~Niu, and X.~Shen, ``Energy-aware
  traffic offloading for green heterogeneous networks,'' \emph{IEEE Journal on
  Selected Areas in Communications}, vol.~34, no.~5, pp. 1116--1129, May 2016.

\bibitem{ra2}
Z.~Kuang, L.~Zhang, and L.~Zhao, ``Energy- and spectral-efficiency tradeoff
  with $\alpha$-fairness in energy harvesting {D2D} communication,'' \emph{IEEE
  Transactions on Vehicular Technology}, vol.~69, no.~9, pp. 9972--9983, Sept.
  2020.

\bibitem{ra3}
F.~Wang, J.~Xu, and S.~Cui, ``Optimal energy allocation and task offloading
  policy for wireless powered mobile edge computing systems,'' \emph{IEEE
  Transactions on Wireless Communications}, vol.~19, no.~4, pp. 2443--2459,
  Apr. 2020.

\bibitem{ra4}
A.~Shahid, V.~Maglogiannis, I.~Ahmed, K.~S. Kim, E.~De~Poorter, and I.~Moerman,
  ``Energy-efficient resource allocation for ultra-dense licensed and
  unlicensed dual-access small cell networks,'' \emph{IEEE Transactions on
  Mobile Computing}, vol.~20, no.~3, pp. 983--1000, Mar. 2021.

\bibitem{eh4}
J.~Huang, B.~Yu, C.-C. Xing, T.~Cerny, and Z.~Ning, ``Online energy scheduling
  policies in energy harvesting enabled {D2D} communications,'' \emph{IEEE
  Transactions on Industrial Informatics}, vol.~17, no.~8, pp. 5678--5687, Aug.
  2021.

\bibitem{ra5}
X.~Wang, Y.~Zhang, R.~Shen, Y.~Xu, and F.-C. Zheng, ``{DRL}-based
  energy-efficient resource allocation frameworks for uplink {NOMA} systems,''
  \emph{IEEE Internet of Things Journal}, vol.~7, no.~8, pp. 7279--7294, Aug.
  2020.

\bibitem{ra6}
P.~Yu, M.~Yang, A.~Xiong, Y.~Ding, W.~Li, X.~Qiu, L.~Meng, M.~Kadoch, and
  M.~Cheriet, ``Intelligent-driven green resource allocation for industrial
  internet of things in {5G} heterogeneous networks,'' \emph{IEEE Transactions
  on Industrial Informatics}, vol.~18, no.~1, pp. 520--530, Jan. 2022.

\bibitem{related2}
S.~Kim and B.~Shim, ``Energy-efficient millimeter-wave cell-free systems under
  limited feedback,'' \emph{IEEE Transactions on Communications}, vol.~69,
  no.~6, pp. 4067--4082, Jun. 2021.

\bibitem{related}
B.~Li, Y.~Dai, Z.~Dong, E.~Panayirci, H.~Jiang, and H.~Jiang,
  ``Energy-efficient resources allocation with millimeter-wave massive {MIMO}
  in ultra dense {HetNets} by {SWIPT} and {CoMP},'' \emph{IEEE Transactions on
  Wireless Communications}, vol.~20, no.~7, pp. 4435--4451, July 2021.

\bibitem{32}
Z.~{Wang}, Q.~{Liu}, M.~{Li}, and W.~{Kellerer}, ``Energy efficient analog
  beamformer design for {mmWave} multicast transmission,'' \emph{IEEE
  Transactions on Green Communications and Networking}, vol.~3, no.~2, pp.
  552--564, Jun. 2019.

\bibitem{31}
J.~{Zhang}, Y.~{Huang}, J.~{Wang}, R.~{Schober}, and L.~{Yang},
  ``Power-efficient beam designs for millimeter wave communication systems,''
  \emph{IEEE Transactions on Wireless Communications}, vol.~19, no.~2, pp.
  1265--1279, Feb. 2020.

\bibitem{3GPP}
J.~D. Oliva~Sánchez and J.~I. Alonso, ``A two-hop {MIMO} relay architecture
  using {LTE} and millimeter wave bands in high-speed trains,'' \emph{IEEE
  Transactions on Vehicular Technology}, vol.~68, no.~3, pp. 2052--2065, Mar.
  2019.

\bibitem{cn}
G.~Noh, B.~Hui, and I.~Kim, ``High speed train communications in {5G}: Design
  elements to mitigate the impact of very high mobility,'' \emph{IEEE Wireless
  Communications}, vol.~27, no.~6, pp. 98--106, Dec. 2020.

\bibitem{28}
L.~{Yan}, X.~{Fang}, L.~{Hao}, and Y.~{Fang}, ``A fast beam alignment scheme
  for dual-band {HSR} wireless networks,'' \emph{IEEE Transactions on Vehicular
  Technology}, vol.~69, no.~4, pp. 3968--3979, Apr. 2020.

\bibitem{29}
K.~{Xu}, Z.~{Shen}, Y.~{Wang}, and X.~{Xia}, ``Location-aided {mMIMO} channel
  tracking and hybrid beamforming for high-speed railway communications: An
  angle-domain approach,'' \emph{IEEE Systems Journal}, vol.~14, no.~1, pp.
  93--104, Mar. 2020.

\bibitem{channel}
J.~Bian, C.-X. Wang, X.~Gao, X.~You, and M.~Zhang, ``A general {3D}
  non-stationary wireless channel model for {5G} and beyond,'' \emph{IEEE
  Transactions on Wireless Communications}, vol.~20, no.~5, pp. 3211--3224, May
  2021.

\bibitem{handover}
R.~Ma, J.~Cao, D.~Feng, H.~Li, and S.~He, ``{FTGPHA}: Fixed-trajectory group
  pre-handover authentication mechanism for mobile relays in {5G} high-speed
  rail networks,'' \emph{IEEE Transactions on Vehicular Technology}, vol.~69,
  no.~2, pp. 2126--2140, Feb. 2020.

\bibitem{zhangjiayi}
J.~Zhang, H.~Du, P.~Zhang, J.~Cheng, and L.~Yang, ``Performance analysis of
  {5G} mobile relay systems for high-speed trains,'' \emph{IEEE Journal on
  Selected Areas in Communications}, vol.~38, no.~12, pp. 2760--2772, Dec.
  2020.

\bibitem{mu}
L.~{Yan}, X.~{Fang}, and C.~{Wang}, ``Position-based limited feedback scheme
  for railway {MU-MIMO} systems,'' \emph{IEEE Transactions on Vehicular
  Technology}, vol.~65, no.~10, pp. 8361--8370, Oct. 2016.

\bibitem{gao2020Efficient}
M.~{Gao}, B.~{Ai}, Y.~{Niu}, W.~{Wu}, P.~{Yang}, F.~{Lyu}, and X.~{Shen},
  ``Efficient hybrid beamforming with anti-blockage design for high-speed
  railway communications,'' \emph{IEEE Transactions on Vehicular Technology},
  vol.~69, no.~9, pp. 9643--9655, Sept. 2020.

\bibitem{justification_ML}
T.~Kim, K.~Ko, I.~Hwang, D.~Hong, S.~Choi, and H.~Wang, ``{RSRP}-based doppler
  shift estimator using machine learning in high-speed train systems,''
  \emph{IEEE Transactions on Vehicular Technology}, vol.~70, no.~1, pp.
  371--380, Jan. 2021.

\bibitem{34}
V.~{Va}, X.~{Zhang}, and R.~W. {Heath}, ``Beam switching for millimeter wave
  communication to support high speed trains,'' in \emph{Proc. IEEE
  VTC2015-Fall}, Boston, MA, Sept. 2015, pp. 1--5.

\bibitem{lei}
L.~{Wang}, B.~{Ai}, Y.~{Niu}, X.~{Chen}, and P.~{Hui}, ``Energy-efficient power
  control of train–ground mmwave communication for high-speed trains,''
  \emph{IEEE Transactions on Vehicular Technology}, vol.~68, no.~8, pp.
  7704--7714, Aug. 2019.

\bibitem{35}
{Qian Chen}, {Xiaoming Peng}, {Juan Yang}, and F.~{Chin}, ``Spatial reuse
  strategy in {mmWave} {WPANs} with directional antennas,'' in \emph{Proc. IEEE
  GLOBECOM 2012}, Anaheim, CA, Dec. 2012, pp. 5392--5397.

\end{thebibliography}

\end{document}